\documentclass[10pt,journal,compsoc]{IEEEtran}

\ifCLASSOPTIONcompsoc
  \usepackage[nocompress]{cite}
\else
  \usepackage{cite}
\fi

\usepackage[pdftex]{graphicx}
\usepackage{amsmath}
\ifCLASSOPTIONcompsoc
  \usepackage[caption=false,font=footnotesize,labelfont=sf,textfont=sf]{subfig}
\else
  \usepackage[caption=false,font=footnotesize]{subfig}tse-cs
\fi

\usepackage{url}
\usepackage{tabularx}
\usepackage{booktabs}
\usepackage{mdframed}
\usepackage{tikz}
\usetikzlibrary{positioning}

\begin{document}

\newcommand{\FSCORE}{\textit{F1 score}}
\newcommand{\RECALL}{\textit{recall}}
\newcommand{\PRECISION}{\textit{precision}}
\newcommand{\etal}{~\textit{et al.}}

\title{On the validity of pre-trained transformers for natural language processing in the software engineering domain}

%
%
%
%

\author{Julian von der Mosel,
        Alexander Trautsch,
        and~Steffen Herbold
\IEEEcompsocitemizethanks{\IEEEcompsocthanksitem J. von der Mosel and A. Trautsch were with the Institute 
of Computer Science, University of Goettingen, Germany. 
GA, 30332.\protect\\
E-mail: alexander.trautsch@cs.uni-goettingen.de
\IEEEcompsocthanksitem S. Herbold was with the Institute of Software and System Engineering, TU Clausthal, Germany.\protect\\
E-mail: steffen.herbold@tu-clausthal.de}
\thanks{Manuscript received XXX, 2021; revised XXX.}}

%
%

\markboth{Journal of \LaTeX\ Class Files,~Vol.~14, No.~8, August~2015}%
{Shell \MakeLowercase{\textit{et al.}}: Bare Demo of IEEEtran.cls for Computer Society Journals}
%



\IEEEtitleabstractindextext{%
\begin{abstract}
Transformers are the current state-of-the-art of natural language processing in many domains and are using traction within software engineering research as well. Such models are pre-trained on large amounts of data, usually from the general domain. However, we only have a limited understanding regarding the validity of transformers within the software engineering domain, i.e., how good such models are at understanding words and sentences within a software engineering context and how this improves the state-of-the-art. Within this article, we shed light on this complex, but crucial issue. We compare BERT transformer models trained with software engineering data with transformers based on general domain data in multiple dimensions: their vocabulary, their ability to understand which words are missing, and their performance in classification tasks. Our results show that for tasks that require understanding of the software engineering context, pre-training with software engineering data is valuable, while general domain models are sufficient for general language understanding, also within the software engineering domain. 
\end{abstract}

}

\maketitle

\IEEEdisplaynontitleabstractindextext

%
\IEEEpeerreviewmaketitle

\IEEEraisesectionheading{\section{Introduction}}
\label{sec:introduction}

\IEEEPARstart{T}{he} introduction of the transformer model \cite{DBLP:journals/corr/VaswaniSPUJGKP17} has permanently changed the field of Natural Language Processing (NLP) and paved the way for modern language representation models such as BERT \cite{devlin-etal-2019-bert}, XLNet \cite{xlnet}, and GPT-2 \cite{radford2019language}. These models have in common that they use transfer learning in the form of pre-training to learn a general representation of language, which can then be fine-tuned to various downstream tasks. Pre-training is expensive and requires large text corpora. Therefore, most of the available models are pre-trained by large companies on vast amounts of general domain data such as the entire English Wikipedia or the Common Crawl News dataset \cite{nagel2016}.
While these models achieve remarkable results in a variety of NLP tasks, the learned word representations (embeddings) still reflect the general domain. This is a problem because the meaning of words varies based on context and is therefore domain dependent. Hence, there is a considerable interest in adapting language representation models to different domains. For example, SciBERT \cite{SciBERT}, BioBERT \cite{BioBERT} and ClinicalBERT \cite{ClinicalBERT} are BERT models adopted to the bio-medical domain. 

In Software Engineering (SE) there are many technical terms that do not exist in other domains and words that have a different meaning within the domain. For example, the word ``bug'' in the general domain refers to an insect, but within the SE domain it refers to a defect. Similarly, ``ant'' is an insect in the general domain, but a build tool within SE. Words such as ``"bug'' and ``ant'' are called polysemes, i.e., words that have different meanings depending on their context. The notion that pre-trained natural language models should be adopted for the SE domain is not new and was already considered by other, e.g., for word2vec embeddings~\cite{Efstathiou2018} and Named Entity Recognition (NER) with BERT models \cite{tabassum-etal-2020-code}.\footnote{Pre-trained models for source code are outside of our scope.}

A gap within the previous research is the larger context. The work by Efstathiou\etal~\cite{Efstathiou2018} only established a difference for the by now dated word2vec approach and not for transformer models. Moreover, the focus is solely on the position of a small set of terms within a vector space. This ignores the ability of transformer models to account for the context, which may be able to infer the correct meaning of polysemes. And while Tabassum\etal~\cite{tabassum-etal-2020-code} successfully demonstrated that pre-training with SE data is also useful with transformer models, they used a small variant of the BERT model and did not explore why the model yielded better results. Thus, it is unclear how much better such a model is at capturing the correct meaning of words and if larger transformer models, as they are usually used within the state-of-the-art of NLP, would perform. 

Within this work, we want to close this gap through an exploratory study that aims to answer the following research questions.

\begin{description}
\item[\textbf{RQ1:}] Are large pre-trained NLP models for SE able to capture the meaning of SE vocabulary correctly?
\item[\textbf{RQ2:}] Do large pre-trained NLP models for SE outperform models pre-trained on general domain data and smaller NLP models that do not require pre-training in SE applications?
\end{description}

We study these research questions using \textit{seBERT}, a BERT\textsubscript{LARGE} model we pre-trained on textual data from Stack Overflow, GitHub, and Jira Issues and BERToverflow, a BERT\textsubscript{BASE} model that was pre-trained by Tabassum\etal~\cite{tabassum-etal-2020-code} on data from Stack Overflow. 

For the first research question, we evaluate the capability of the models to correctly determine the meaning of terms in three ways: the comparison of the vocabularies of the models; their capability to infer missing words in a list of curated sentences; and the capabilities to infer missing polysemes both in a SE domain corpus and a general domain corpus. Inferring missing words is a task typically referred to as Masked Language Modeling (MLM). Through this, we complement the study by Efstathiou\etal~\cite{Efstathiou2018} to understand the ability of NLP models for SE to not only create valid embeddings, but even predict the correct words given the context. Such a prediction goes beyond similarity of single words and would be a strong indicator that NLP models for the SE domain should outperform general domain models because they are demonstrably better at capturing the correct meaning. 

For the second research question, we study the capability of these models to improve the performance of prediction tasks. We use several classification tasks for this purpose. 1)~The prediction of bug issues, similar to Herbold\etal~\cite{Herbold2020}. This task allows us to study how well the models perform with relatively long text for a task that is, in fact, often performed inaccurately by humans~\cite{Herzig2013}. 2)  The identification of quality improving commits~\cite{Trautsch2021}, i.e., a task with very short texts. 3) Sentiment mining, a tasks where it is was already shown that general domain transformers perform well~\cite{Zhang2020}.

Through our study, we provide the following contributions to the literature. 

\begin{itemize}
\item An analysis of the validity of NLP models for SE through a combination of comparison of vocabularies and the prediction of missing words. We show that BERT models trained with SE data are able to capture the correct meaning of SE terminology, at the cost of the correct modeling of some general domain concepts like geographical locations. We further show that larger models trained with more diverse data are better at capturing nuanced meanings than smaller domain specific models. 
\item We advance the state of issue type prediction and commit intent prediction by improving the model performance significantly with the fine-tuned SE specific BERToverflow and seBERT models. Our data indicates that the improvement is only possibly because the model was pre-trained on SE data and also that larger NLP models outperform smaller models in the SE domain similar to the differences that can be observed between general domain models.
\item We found that sentiment mining was not strongly affected by domain specific pre-training and that SE domain transformers perform similar to general domain transformers for this task. 
\item A basic ethical evaluation of the models revealed that the SE domain models should only be used with care, because they may to have some troubling properties, especially with respect to gender bias. This warrants further research and until these aspects are better understood, such models should not be used for any tasks that may be negatively affected by gender bias, though this also warrants caution regarding other biases such as racial bias. 
\end{itemize}

The remainder of this article is structured as follows. We provide some background on transformers in Section~\ref{sec:background}, followed by a discussion of related work on domain specific NLP models in Section~\ref{sec:related-work}. Then, we outline the creation of the SE domain model seBERT in Section~\ref{sec:sebert}. Based on these foundations, we introduce our method for validating transformers within the SE domain in Section~\ref{sec:validation}, present the results of this validation in Section~\ref{sec:results}, and discuss these results in Section~\ref{sec:discussion}. Finally, we conclude in Section~\ref{sec:conclusion}.

\section{Background on Transformers}
\label{sec:background}

NLP is a broad field composed of various disciplines such as computational linguistics, machine learning, artificial intelligence, computer science, and speech processing \cite{eisenstein}. Modern NLP models have a variety of applications, including text classification, NER, machine translation, sentiment analysis, and natural language generation. In recent years, transformer models, e.g., BERT~\cite{devlin-etal-2019-bert}, BART~\cite{Lewis2020}, ALBERT~\cite{Lan2020ALBERT}, RoBERTa~\cite{liu2019roberta}, GPT-3~\cite{brown2020language}, and Switch-C~\cite{fedus2021switch} established themselves at the state-of-the-art of NLP within the general domain. Since the focus of our article is on the validity of transformer models within the SE domain, we avoid an in-depth technical discussion of the underlying neural network architectures of these models and refer for this to the literature instead~(e.g., \cite{Tunstall2021}). Instead, we outline the concept of such models on a high level in natural language without mathematical details. 

\begin{figure}
\centering
\includegraphics[width=.4\textwidth]{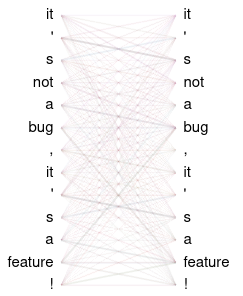}
\caption{Visualization of self-attention that shows how the different words in the context on the left side influence the meaning of the words on the right side. The figure shows the attention of the first transformer layer of seBERT calculated with BertViz \cite{Vig2019}.}
\label{fig:self-attention}
\end{figure}

Transformers can be described in a single sentence: they are \textit{sequence-to-sequence} networks with \textit{self-attention}. Now we only need to understand what the highlighted aspects mean. A sequence-to-sequence network takes as input a sequence and generates a new sequence of the same length. For NLP, the input sequence is usually a tokenized text, possibly augmented with additional tokens, e.g., to represent classes. The input sequence is then encoded, such that each input token is represented by a numeric vector, which means that we have a sequence of vectors through these embeddings. This transformation relies on the self-attention, which means that the tokens are not considered in isolation, but together. Consider this paragraph: the self-attention would consider this whole paragraph at once and, for each word, model how the \textit{contextual meaning} depends on all other words in the paragraph. This is shown in Figure~\ref{fig:self-attention}. The figure shows, e.g., that the meaning of ``bug'' is strongly influenced by the words ``not'' and ``feature'', and that the meaning of ``s'' is defined by the apostrophe. Thus, there is no fixed meaning of each word and the meaning depends on the context: ``bug'' actually means ``not bug'' or ``feature'', ``s'' is only a contraction of ``is''.

That the whole sequence is used as context for each word to understand the contextual meaning is the major difference to previously used NLP models. For example, word embeddings organize words such that words with similar meanings are close to each other~\cite{mikolov2013efficient}. However, while the context is taken into account when the embedding is calculated, each word only gets a single position. If words are polysemes, the meaning depends on the context, which cannot be accurately encoded with such a fixed embedding. Either one contextual meaning is lost, e.g., ``bug'' is only close to other insects, but not to words like ``defect''. Or different concepts become similar to each other, e.g., ``bug'' is close to both insects and defects, which means that insects in general are now similar to defects. Moreover, since the embedding does not take the position of a word in an actual sentence into account, the interpretation of sentences based on the grammatical relationship between words is also not possible. Recurrent Neural Networks (RNN) without self-attention, such as Long-Short-Term-Memory (LSTM) networks, can also not take the complete context accurately into account. While such models can also be sequence-to-sequence networks, the relationship between a word at a specific position, and all other words, is linear. Simplified, this means that the influence on the context depends mostly on the distance to a specific word. In comparison, the self-attention learns how each word influences the context, depending on the position without a fixed influence of the distance. For our example from Figure~\ref{fig:self-attention}, this means that the long distance between the first ``it's'' and the last word ``feature'' does not mean that there cannot be a strong direct relationship.

Due to these differences, transformers quickly became the most powerful architecture for neural network based NLP. Models based on transformers are the current state of the art. The revolution was spear-headed by BERT~\cite{devlin-etal-2019-bert}, followed by similar models such as RoBERTa~\cite{liu2019roberta}, GPT-3~\cite{brown2020language} that were pre-trained with more data, used larger neural networks, and/or optimized the efficiency of the self-attention mechanisms. Such models already gained traction within the SE domain: Zhang\etal~\cite{Zhang2020}, who recently established transformers as state-of-the-art for sentiment mining within the SE domain, clearly outperforming smaller sentiment mining models like SentiStrength~\cite{Thelwall2010}, including models tailored specifically to the SE domain like SentiStrengh-SE~\cite{ISLAM2018125}, SentiCR~\cite{Ahmed2017}, and Senti4SD~\cite{Calefato2018}. Biswas\etal~\cite{Biswas2020} also found a strong performance of BERT for sentiment mining. 

The drawback of transformer models is the size of the neural networks. Even a relatively small model like BERT\textsubscript{BASE}~\cite{devlin-etal-2019-bert} already has 110 million parameters. The largest currently discussed transformer architectures like GPT-3~\cite{brown2020language} and Switch-C~\cite{fedus2021switch} have more than 100 billion parameters. Thus, the training of such models requires dedicated hardware and huge amounts of data. To reduce the burden, these models are \textit{pre-trained} in a \textit{self-supervised} setting. This means that the neural networks are trained on a large corpus of NLP data, such that the structure of text and meaning of words is known by the network. Self-supervised means that this understanding of the language is trained in a supervised way (i.e., with labels), but that the labels are generated directly from the training data. We will show how this works for the BERT pre-training in Section~\ref{sec:pretrain}. Researchers and practitioners who then want to apply NLP to solve a problem, use these pre-trained models and \textit{fine-tune} the models for a specific task on a labeled data set. This works with less data and requires only few training steps and is, therefore, usually not computationally expensive. However, this may still require dedicated hardware, due to the size of the models. 

\section{Related Work on Domain-Specific NLP}
\label{sec:related-work}

In recent years there has been considerable interest in adapting pre-trained language models such as word2vec~\cite{mikolov2013efficient}, ELMo~\cite{peters-etal-2018-deep} and BERT to different domains. However, most of the related work concerns the biomedical or financial domain. Pyysalo\etal~\cite{Pyysalo2013} were among the first to provide domain-specific word2vec embeddings based on biomedical corpora. Since then, recent biomedical adaptions of word2vec and BERT include Dis2Vec \cite{Dis2Vec}, BioBERT \cite{BioBERT}, ClinicalBERT \cite{ClinicalBERT} and to some extent SciBERT~\cite{SciBERT}, which has been pre-trained on biomedical and computer science papers. Models such as BioBERT and SciBERT have been shown to outperform BERT in biomedical tasks and achieve state-of-the-art results \cite{SciBERT,BioBERT}. The same is true for FinBERT~\cite{FinBERT} which achieved state-of-the-art results in financial sentiment analysis. However, except for SciBERT, the domain-specific BERT models are not pre-trained from scratch, but use the same vocabulary as BERT or are further pre-trained using BERTs weights. 

Efstathiou\etal~\cite{Efstathiou2018} pre-trained a general-purpose word representation model for the software engineering domain. They trained the so\_word2vec on 15GB of textual data from Stack Overflow posts. The authors compare the so\_word2vec model with the original Google news word2vec model and show that it performs well in capturing SE specific meanings and analogies. The main difference with our work is that we consider contextual embeddings and provide a more extensive evaluation of the validity of the resulting embeddings. In addition, our seBERT model was trained using 119GB of data from multiple sources, i.e., we also use GitHub issues and commit messages in addition to data from Stack Overflow. Ferrari and Esuli~\cite{Ferrari2019} used word embeddings to identify terms that are shared between domains, but which may have different meanings. While their work was not on the difference between the SE domain and the general domain, they showed that word embeddings trained on corpora from different domains indeed yield different embeddings for polysemous word, which provides another indication that general domain models may have problems when used for domain-specific purposes.

Recently, Tabassum\etal~\cite{tabassum-etal-2020-code} proposed a similar approach to ours, in which GloVe, ELMo, and BERT models are pre-trained on 152M sentences from Stack Overflow. In addition, the authors propose a novel attention based SoftNER model designed for code and NER. Their BERT model BERToverflow was pre-trained using a cased 64000 WordPiece vocabulary with the same configuration as original BERT\textsubscript{BASE} with 110 million parameters. The results show that BERToverflow clearly outperforms the other models, inluding BERT\textsubscript{BASE}, on NER tasks. In contrast, we pre-train seBERT with about six times more data, including data from GitHub and Jira. Morover, seBERT uses an uncased 30522 WordPiece vocabulary and the same configuration as BERT\textsubscript{LARGE}, resulting in a much larger model with 340 million parameters. Additionally, our focus is different and not on NER tasks, but rather on the general validity of the NLP models and their usefulness for classification tasks in a pure NLP setting without considering source code. 

While our focus is on pure NLP, we also want to mention similar models from the SE domain for code. Theeten\etal~\cite{Theeten2019} proposed import2vec, a word2vec based approach for learning embeddings for software libraries. The embeddings are trained on import statements extracted from source code of Java, JavaScript and Python open source repositories. In their work, they show that their embeddings capture aspects of semantic similarity between libraries and that they can be clustered by specific domains or platform libraries. A more general source code based pre-training was conducted by Alon\etal~\cite{Alon2019} who proposed code2vec to represent code snippets as word embeddings. The authors show that such embeddings are a powerful tool to predict method names based on the code. Another source code based approach is CodeBERT \cite{feng2020codebert}, a bimodal pre-trained bidirectional transformer for natural language and programming languages. CodeBERT is pre-trained using a hybrid objective function combining the default MLM task and a Replaced Token Detection (RTD) \cite{clark2020electra} task. For the training data, the authors use both bimodal data pairs comprising a function (code) and its documentation (natural language), as well as unimodal data consisting of only of the function. The data was collected from GitHub code repositories in six different programming languages. Their results show that CodeBERT achieves state-of-the-art performance in natural language code search and code-to-document generation tasks. 

\section{seBERT}
\label{sec:sebert}

Within this section we describe how we trained seBERT, a large-scale transformer model for the SE domain that complements BERToverflow, because it is larger, was trained on more data diverse data with a larger volume, and is uncased (see Section~\ref{sec:related-work}).

\subsection{Data}
\label{sec:data}

We identified four data sources for our domain-specific corpus of SE textual data: 

\begin{enumerate}
\item \textbf{Stack Overflow posts}: With millions of questions, answers and comments, Stack Overflow posts are a rich source of textual data from the SE field. Same as the prior work, this is one of our main data sources. The Stack Exchange Data Dump \cite{stack-exchange-data-dump} contains Stack Overflow posts from 2014 to 2020 and is hosted on the Internet Archive \cite{internet-archive}. A Stack Overflow specific mirror is available as a public dataset on the Google Cloud Platform \cite{stackoverflow-dataset}. Using BigQuery, we extracted 62.8 Gigabyte of questions, answers and comments.
\item \textbf{GitHub issues}: GitHub is more than a code hosting environment. For many users, GitHub issues are the first place to give feedback or report software bugs and thus provide valuable insight into the communication between users and developers. GitHub issues consist of a title, a description and comments and are available through the GitHub Archive \cite{gh-archive}. Using Google BigQuery \cite{google-bigquery} we extracted 118.5 Gigabyte of issue descriptions and comments from the years 2015 to 2019.
\item \textbf{Jira issues}: Similar to GitHub issues, Jira issues provide valuable insights into software team communication regarding bug and issue tracking. In 2015, Ortu et al. published the Jira Repository Dataset \cite{the-jira-repository-dataset}, which contains 700K Jira issue reports and more than 2M Jira issue comments. With 1.4 Gigabyte of unprocessed textual data, they make up the smallest part of our corpus. However, the overall amount of data is still fairly large, and provides a perspective beyond Stack Overflow and GitHub. This includes language regarding the committing solutions to Subversion, or the submission of textual diffs as patches.
\item \textbf{GitHub commit messages}: Git commit messages are used to communicate the context of changes to other developers. Commit messages consist of two parts, a subject line and a message body, the latter being optional. Similar to the GitHub issues, we extracted 21.7 Gigabyte of GitHub commit messages from the GitHub Archive using BigQuery.
\end{enumerate}

In total, our data set consists of 204.4 Gigabyte of unprocessed textual data. To the best of our knowledge, this the largest and most diverse corpus of textual data in the SE domain. Prior work on pre-training focused only on data from Stack Overflow. However, this ignores important aspects of the SE domain. A key aspect of natural language communication is the description of feature requests and bugs. Such data is typically not available on Stack Overflow, with the exception of users asking how they could work around a specific issue. Through the inclusion of GitHub issues and Jira issues, we enhance the corpus with such data. Moreover, commit messages are often short and on point natural language summaries of development activities and, consequently, different from the often longer discussions within issues and on Stack Overflow.

\subsection{Preprocessing}
\label{sec:preprocessing}

The textual data is mostly unstructured and needs to be preprocessed. Overall, we conducted eight different preprocessing steps. 
\begin{enumerate}
    \item \textbf{Basic preprocessing}: We convert all documents to lowercase, remove control characters (newline, carriage return, tab) and normalize quotation marks and whitespaces.
    \item \textbf{English}: Since we are training a model for the English language, we use the fastText\footnote{https://fasttext.cc/} library to remove all non-English documents.
    \item \textbf{HTML}: We remove HTML tags and extract text using the BeautifulSoup library \footnote{https://beautiful-soup-4.readthedocs.io/}.
    \item \textbf{Markdown}: We use regular expressions to greedily remove Markdown formatting.
    \item \textbf{Hashes}: Hashes such as SHA-1 or md5 do not provide any contextual information and should be removed. We detect hashes by checking whether alphanumeric words with a length of 7 characters or more can be cast to a hexadecimal number and replace them with [HASH] tokens.
    \item \textbf{Code}: Source code is not natural language and should be removed. However, finding code fragments within text is a non-trivial task. We use HTML \textless{}code\textgreater{} tags, Markdown code blocks and other formatting to identify source code and replace it with [CODE] tokens. Code that is not within such environments is not filtered. 
    \item \textbf{User mentions}: For privacy reasons, we replace usernames and mentions (@user) with [USER] tokens.
    \item \textbf{Special formatting}: We remove special formatting and content such as Jira specific formatting, Git sign-off, or references to SVN.
\end{enumerate}

The preprocessing steps differ for the data sources. For example, removal of Markdown does not make sense for Stack Overflow posts or commit messages. Table~\ref{tab:preprocessing} shows which steps we applied to which data and Table~\ref{tab:data} shows the amount of data from each data source after preprocessing. After preprocessing, a total of 119.7 Gigabyte of text or 20.9 billion words remain.

\begin{table*}[t]
\centering
\begin{tabular}{lcccc}
\toprule
\textbf{Processing} & \textbf{GitHub issues} & \textbf{Commit messages} & \textbf{Stack Overflow} & \textbf{Jira issues}  \\
\midrule
Basic &  X  & X & X & X \\
English & X & X &  &  \\
HTML &  &  & X &  \\
Markdown & X &  &  & \\ 
Hashes & X & X & X & X \\ 
Code & X &  & X & X \\
User mentions & X &  & X & \\
Special formatting &  & X &  & X  \\ 
\bottomrule
\end{tabular}
\caption{The applied preprocessing steps for each data source.}
\label{tab:preprocessing}
\end{table*}

\begin{table*}[t]
\centering
\begin{tabular}{lll}
\toprule
\multicolumn{1}{c}{\textbf{GitHub}} &  \multicolumn{1}{c}{\textbf{Stack Overflow}} & \multicolumn{1}{c}{\textbf{Jira}} \\ 
\midrule
\begin{tabular}[t]{@{}l r@{}} \textbf{Issues} &  29.7 Gigabyte \\ \textbf{Comments} & 39.3 Gigabyte \\ \textbf{Commit messages} & 18.2 Gigabyte  \end{tabular} 
& \begin{tabular}[t]{@{}l r@{}} \textbf{Questions} &  10 Gigabyte \\ \textbf{Answers} & 10.1 Gigabyte\\ \textbf{Comments} & 11.3 Gigabyte  \end{tabular} 
& \begin{tabular}[t]{@{}l r@{}} \textbf{Issues} &  502 Megabyte \\ \textbf{Comments} & 613 Megabyte   \end{tabular} \\ \bottomrule 
\end{tabular}
\caption{The size of the data after preprocessing.}
\label{tab:data}
\end{table*}

\subsection{Pre-training}
\label{sec:pretrain}

The BERT implementation by Devlin\etal~\cite{devlin-etal-2019-bert} uses WordPiece embeddings with a 30522 token vocabulary. We train a new SE domain-specific vocabulary of the same size using all our data and the BertWordPieceTokenizer by HuggingFace.\footnote{https://huggingface.co/}

\begin{figure*}[t]
\centering
\includegraphics[width=0.32\textwidth]{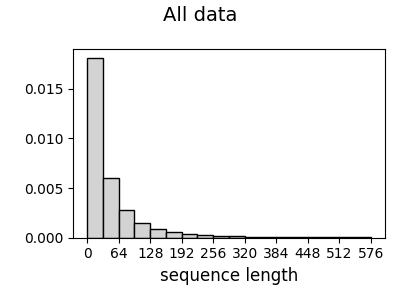}
\includegraphics[width=0.32\textwidth]{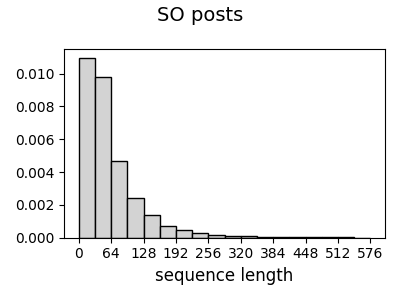}

\includegraphics[width=0.32\textwidth]{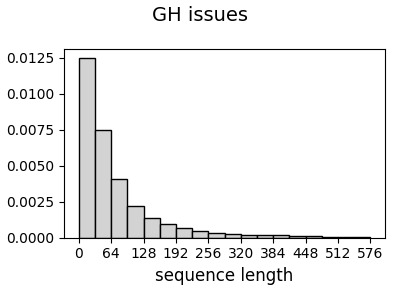}
\includegraphics[width=0.32\textwidth]{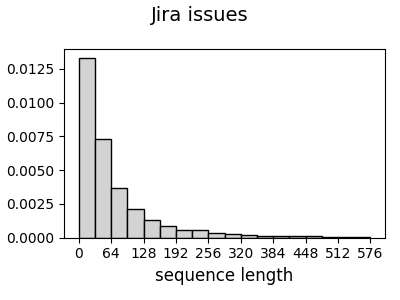}
\includegraphics[width=0.32\textwidth]{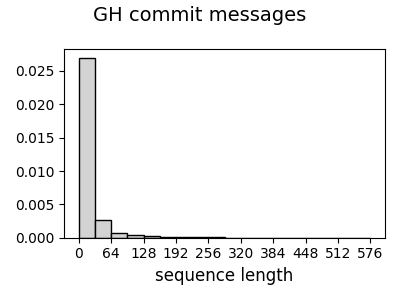}
\caption{Histograms of sequence length for each data source and all data combined.}
\label{fig:hists}
\end{figure*}

An important parameter in BERT pre-training is the maximum sequence length. Training sequences shorter than the maximum sequence length are padded with [PAD] tokens, while longer sequences are truncated. Since the self-attention mechanism of BERT has a quadratic complexity with respect to the sequence length~\cite{devlin-etal-2019-bert}, the parameter significantly affects the training time and the memory requirements. Therefore, the authors recommend training 90\% of the training steps with a sequence length of 128 and 10\% of the steps with a sequence length of 512 to learn longer contexts \cite{devlin-etal-2019-bert}. We analyzed the sequence length for all our data through histograms, which are shown in Figure \ref{fig:hists}. The sequence lengths follow exponential distribution and 96.7\% of all training sequences are shorter than 256 and 90.1\% are shorter than 128. Therefore, since most of our data is short-sequence data, we train with a sequence length of 128 for all steps at the cost of a possible small disadvantage with very long contexts.

BERT pre-training is self-supervised based on MLM and Next Sentence Prediction (NSP) tasks. For the NSP task, we format the data using the NLTK library\footnote{https://www.nltk.org/} so that each line contains a sentence and documents are separated by blank lines. For the NSP task, the input data is then prepared such that each subsequent pairs of sentences makes up one input sequence of the form \texttt{[CLS] sentence\_1 [SEP] sentence\_2 [SEP]}. We re-use code from the original BERT to prepare the data for the MLM task.\footnote{https://github.com/google-research/bert}  The provided script duplicates the input sequences by a dupe factor and creates training samples by randomly masking 15\% of the whole words. Whole word masking was not part of the original BERT implementation and only added later by the authors as improvement of the preprocessing. With whole word masking all tokens of a word are masked at once, which means that it is not possible that words are only partially masked. Figure~\ref{fig:masking} shows this for a document with three sentences. As a result of this data preparation, we have 2.4 Terabyte of data which we can use as input for the pre-training of BERT with TensorFlow.

\begin{figure}
\tiny
\begin{tikzpicture}
    \node[text width=2cm] at (0,0) {Document with three sentences.};
    \node[text width=2cm] at (0,-1.32) {Two input sequences with sentence pairs.};
    \node[text width=2cm] at (0,-3.65) {Duplicates of the input sequences with 15\% randomly masked words.};
    
    \node[shape=rectangle,draw=lightgray,text width=6.3cm] (t1) at (4.4,0) {i need to run a method every 5 seconds using mono for android.\\is there a scheduled timer in android?\\i have tried this code, however, it fails to start: [CODE]};
    
    \node[shape=rectangle,draw=lightgray,text width=6.3cm,below=0.4cm of t1.south] (t2) {[CLS] i need to run a method every 5 seconds using mono for android\\{} [SEP] is there a scheduled timer in android? [SEP]};
    \node[shape=rectangle,draw=lightgray,text width=6.3cm, below=0.2cm of t2.south] (t3) {[CLS] is there a scheduled timer in android?\\{} [SEP] i have tried this code, however, it fails to start: [CODE] [SEP]};
    
    \node[shape=rectangle,draw=lightgray,text width=6.3cm, below=0.4cm of t3.south] (t4) {[CLS] i need to run a [MASK] every 5 seconds using mono [MASK] android\\{} [SEP] is there a [MASK] timer in android? [SEP]};
    \node[shape=rectangle,draw=lightgray,text width=6.3cm, below=0.2cm of t4.south] (t5) {[CLS] i [MASK] to run a method every 5 seconds using [MASK] for android\\{} [SEP] is there a scheduled [MASK] in android? [SEP]};
    \node[shape=rectangle,draw=lightgray,text width=6.3cm, below=0.2cm of t5.south] (t6) {[CLS] is there a scheduled [MASK] in android?\\{} [SEP] i have tried this [MASK], however, it fails to start: [CODE] [SEP]};
    \node[shape=rectangle,draw=lightgray,text width=6.3cm, below=0.2cm of t6.south] (t7) {[CLS] is there a [MASK] timer in android?\\{} [SEP] i have tried this code, [MASK], it fails to start: [CODE] [SEP]};
    
    \node[shape=rectangle, draw=black, minimum width=8.8cm, minimum height=0.9cm] at (3.3,0) {};
    \node[shape=rectangle, draw=black, minimum width=8.8cm, minimum height=1.5cm] at (3.3,-1.32) {};
    \node[shape=rectangle, draw=black, minimum width=8.8cm, minimum height=2.9cm] at (3.3,-3.65) {};
    
\end{tikzpicture}
\caption{Example for preparation of data for training with BERT.}
\label{fig:masking}
\end{figure}


We pre-train seBERT using the same configuration as BERT\textsubscript{LARGE}, i.e., with 24 layers, a hidden layer size of 1024, and 16 self-attention heads, which leads to a total of 340 million parameters. Pre-training of BERT is expensive, e.g., the original BERT\textsubscript{LARGE} was pre-trained on 16 Cloud TPUs for 3 days. Since then, several optimizations have been proposed to reduce the training time of BERT. You et al.~\cite{LAMB} proposed a Layer-wise Adaptive Moments Based (LAMB) optimizer that allows the use of larger minibatches compared to the originally used ADAMW optimizer. NVIDIA then implemented their own version NVLAMB and combined it with Automatic Mixed Precision (AMP) \cite{DBLP:journals/corr/abs-1710-03740}, which accelerates training time and reduces memory usage by autocasting variables to 16 bit floating point numbers upon retrieval, while storing variables with the usual 32 bit precision.  We used this NVIDIA implementation of the BERT pre-training to facilitate the effective usage of the system we had available with 2x24 CPU cores, 384 GB RAM and 8x NVIDIA Tesla V100 32G GPUs. The pre-training of seBERT required about 4.5 days. 

\section{Validation of SE Domain Models}
\label{sec:validation}

The evaluation of pre-trained language models is usually performed with benchmarks such as the GLUE benchmark~\cite{GLUE}, its successor SuperGLUE~\cite{SuperGLUE} or SQuAD~\cite{SQuAD}. However, these benchmarks require fine-tuning and are designed for the general domain and therefore not suitable for the evaluation of NLP models for the SE domain. Moreover, while the evaluation on benchmark data is useful to evaluate the overall predictive capabilities of a model, they are not suitable to understand the reasoning within a model. Therefore, we propose a different approach for the evaluation of transformer models within the SE domain that considers the validity from three different perspectives: the vocabulary, context sensitive prediction of masked words, and a fine-tuned benchmarks using a limited amounts of data for classification. 

\subsection{Vocabulary}

One of the main differences between a domain-specific language model and a general domain language model should be the vocabulary. We expect that the SE specific models should gain words from the SE domain, while lost words should be from other domains. Since alphanumeric words are more prevalent in the domain, a larger proportion of words should be alphanumeric.

We take pattern from Beltagy et al.~\cite{SciBERT} and perform a vocabulary analysis by computing the WordPiece overlap, i.e., the percentage of identical tokens between vocabularies. Additionally, we examine the vocabularies for differences in structure, tokenization, and the types of words gained or lost through a manual analysis of the differences. The manual analysis was conducted by two authors who both read the word lists and took notes about differences. The notes were then compared and discussed during a virtual meeting, which resulted in the qualitative assessment of the differences we provide in Section~\ref{sec:results-vocab}. 

\subsection{Contextual embeddings}

Transformer models such as BERT go beyond word similarities and instead consider the complete context to infer the meaning. This goes beyond the context-free finding of similar words that was used by Efstathiou\etal~\cite{Efstathiou2018} to determine if word embeddings from the SE domain are valid. Instead, we suggest to use MLM tasks to evaluate the validity of the embeddings, i.e., the evaluation of understanding how good the model is at predicting missing words in a text~\cite{devlin-etal-2019-bert, SciBERT}. We devise three categories of sentences suitable for validating language representation models for the software engineering domain with the goal to understand if seBERT and BERToverflow behave as expected due to the differences in training data from the original BERT. The three categories are the following.  

\begin{enumerate}
    \item \textbf{Positive examples}: Prediction of masked words within a software engineering context, i.e., where the missing word or context words are polysemuous or from the SE domain. We expect domain-specific models to produce accurate predictions, while the original BERT may struggle because of the SE context.
    \item \textbf{Negative examples}: Prediction of masked words outside a software engineering context, where the missing word or context words are polysemous. Given the polysemy of the negative examples, domain-specific models should produce domain-specific predictions, which are not suitable for the sentences from the general domain.
    \item \textbf{Neutral examples}: Prediction of masked words outside a SE context, where the missing word can be inferred from general language understanding (e.g. idioms, opposites, antimetaboles). General language understanding should be preserved across domains, we expect all models to perform well.
\end{enumerate}

Since sentences of these categories can only demonstrate the validity by way of example, we refer to them as \textit{validation examples}. We defined ten examples per category. Due to space restrictions, we only list the sentences together with the predictions by the models in Section~\ref{sec:mlm-results} in tables~\ref{tab:positive}-\ref{tab:neutral}. For each category, we defined ten validation examples. To define these examples, one author created a list of candidate sentences including the word that should be masked and the expectation. These candidates were prepared based on the polysemous terms from Efstathiou\etal~\cite{Efstathiou2018}. Then, all authors discussed and refined these candidates together in a virtual meeting to define the ten examples per category. After this process, we can then compare how the different models complete the sentences and how this matches our expectation. 

The drawback of our validation examples is that while they provide valuable insights, they are not an unbiased large sample, but rather a small and manually designed sample with a specific purpose. This raises potential issues regarding the generalizability of the effects we find to real world data. We define the following experiment with the goal to compare the differences of contextual embeddings within the SE domain on a large scale. The general idea is to conduct a MLM experiment in which polysemes are masked and then compare how good the models are at predicting the polysemes. \begin{itemize}
  \item Determine the overlap between the vocabularies of all involved models and discard all subwords.
  \item Manually extract a list of polysemes from the overlapping words, i.e., words that have a different meaning in the SE domain and the general domain (e.g., shell, bug, root).
  \item Collect two large corpora of text data: one from the SE domain and one from the general domain. Both corpora must not have been used during the pre-training of the models. In our case, we collected the Stack Overflow posts from 2021 as SE domain corpus (2.9 GB) and used the CC-News corpus (2.6 GB)\footnote{\url{https://huggingface.co/datasets/cc\_news}} as general domain corpus.
  \item For each of the polysemes, randomly sample 100 sentences from both corpora. If there are less than 100 such sentences, drop the polysemes from the analysis. 
  \item Mask the polyseme in each sentence and query the model for the five most likely words. 
\end{itemize}
We can then evaluate the different models with respect to their capability to find the masked word. We would expect that models from the SE domain perform better on the Stack Overflow corpus and, vice versa, models from the general domain perform better on the CC-News corpus.

\subsection{Fine-tuned Prediction Tasks}
\label{sec:fine-tune-desc}

The first check regarding the validity is to analyze the pre-trained model directly. However, it is unclear if the actual use of these models in fine-tuned applications would also be affected. We propose to evaluate the validity of the models by applying them to a classification task within the SE domain, for which only a limited amount of data is available. We propose to always use (at least) three different tasks to evaluate the capabilities of the fine-tuning. 
\begin{itemize}
  \item The first type of task should have longer texts. This task is suitable to evaluate how good the model is at understanding longer texts and contexts.
  \item The second type of task should have shorter text, to evaluate how good the model is for short texts.
  \item The third task should be sentiment mining, as a general domain use case that is already well understood in the SE domain~\cite{Novielli2021}, incl. strong, already existing results that general domain transformer models are the current state-of-the-art for sentiment mining in the SE domain~\cite{Zhang2020}. Since the work by Novielli\etal~\cite{Novielli2021} suggests that SE domain models for sentiment mining do not yet work very well because their results are unstable, and Novielli\etal~\cite{Novielli2018}, Lin\etal~\cite{Lin2018}, and Zhang\etal~\cite{Zhang2020} showed that smaller general domain models frequently outperform SE-specific models,\footnote{While this is not the core conclusion of Novielli\etal and Zhang\etal, their results in terms of the class imbalance insensitive macro average show the following: SentiStrength outperforms SentiStrength-SE and SentiCR on the Stack Overflow data by Novielli\etal~\cite{Novielli2018}. SentiStrength outperforms SentiStrength-SE, Senti4SD for the API and SO data, as well as only Senti4SD on the App data by Zhang\etal~\cite{Zhang2020}} this task allows us to evaluate if pre-trained SE domain transformer models are also better on SE domain data for use cases that may not require domain knowledge.
\end{itemize}

For all fine-tuning tasks, we propose to simply re-use the classification implementation for BERT from HuggingFace:\footnote{\url{https://huggingface.co/transformers/v4.2.2/model_doc/bert.html}} this implementation adds a single fully connected layer which produces the classification based on the output sequence of the BERT model. We then train each model for five epochs on the issue, commit, respectively, sentiment data for the fine-tuning. We use 80\% of the training data for the fine-tuning and 20\% as validation data to determine when the models start to overfit. 

We note that we use different fine-tuning tasks than the NER task that was used by Tabassum\etal~\cite{tabassum-etal-2020-code}. Our rationale for using a classification task instead is that the NER task is very specific, especially with the respect that the exploitation of casing and special characters like the underscore is a major aspect of NER within natural language models. Thus, this task does not aim to understand the meaning, but rather the inference of a specific language construct. In comparison, our tasks require the interpretation of the meaning of natural language.

In the following, we discuss how we implement and evaluate each of the fine tuning tasks in detail.

\subsubsection{Prediction of Bug Issues}

We propose to use the prediction if issues are bugs~\cite{Herbold2020} as task with longer sequences. This task has several properties that make it a good candidate to evaluate the validity and benefits of transformer models. Issues are among the longer texts within SE data (see Figure~\ref{fig:hists}), Moreover, there is an established baseline performance which shows that a simple neural network approach without pre-training based on fastText~\cite{fasttext} with built-in automated hyperparameter tuning performs well~\cite{Herbold2020}. Thus, we can not only compare pre-trained models to each other, but also their benefit over simpler models that can be trained in minutes without any special hardware. Second, the task is hard and often not solved correctly by humans, unless careful manual validation is used~\cite{Antoniol2008, Herzig2013, Herbold2020a}.

We use the five projects with validated issue type data from from Herzig\etal~\cite{Herzig2013} together with the 38 validated projects from Herbold\etal~\cite{Herbold2020a} together as a single data set with 43 projects and conduct a leave-one-project-out cross validation experiment. This means that one project is used for testing and the remaining 42 projects for the training of the classifier, i.e., of the a fastText model as baseline and for the fine-tuning of the BERT models. Since there are 38,219 issues in the data and the largest project has 2,399 issues, we create 43 fine-tuned models trained with at least 35,820 issue descriptions. Following Herbold\etal~\cite{Herbold2020}, we conduct a statistical comparison based on the \FSCORE{} which is defined as

\begin{equation*}
\begin{split}
\RECALL &= \frac{tp}{tp+fn} \\
\PRECISION &= \frac{tp}{tp+fp} \\
\FSCORE &= 2 \cdot \frac{recall \cdot
precision}{recall+precision}
\end{split}
\end{equation*}
where $tp$ are true positives, i.e., bugs that are classified as bugs, $tn$ true negatives, i.e., non bugs classified as non bugs, $fp$ bugs not classified as bugs and $fn$ non bugs classified as bugs. Additionally, the \RECALL{} and \PRECISION{} should be reported to allow us to understand the nature of the errors made by the models. We use a Bayesian approach for the statistical comparison following the guidelines by Benavoli\etal~\cite{Benavoli2017}. The authors suggest to use a Bayesian signed rank test for a comparison of results on multiple data sets, as is the case here. The advantage of the Bayesian approach is that the number of models that are compared does not influence the results, as this directly estimates that one approach performs better, which does not require an adjustment of the significance level~\cite{Benavoli2017}. 

\subsubsection{Commit Intent Classification}

As second fine-tuning task with shorter sequences, we use the automated classification of commits based on commit messages. Commit messages are usually very short (see Figure~\ref{fig:hists}). Unfortunately, we are not aware of a solid benchmark for the machine-learning based classification of commit messages. Due to the good results for issues, we assume that fastText is also a good baseline for the classification of commit messages. As data, we use manually validated data from Trautsch\etal~\cite{Trautsch2021}, i.e., we have a sample with a total size of 2533 commit messages that has as class label whether a commit is perfective or not. We follow the approach outlined by Benavoli\etal~\cite{Benavoli2017} for the comparison of classifiers on a single data set: we conduct 10x10 cross-validation and obtain 100 results for each classifier. We then use the Bayesian correlated $t$-test to compare the results. Same as above, we use the \FSCORE{} as foundation for the statistical comparison and report the \RECALL{} and \PRECISION{} to understand the nature of the errors made by the models. 

\subsubsection{Sentiment Mining}

As third fine-tuning task we propose sentiment mining~\cite{Novielli2021, Zhang2020}. We use three data sets that were also used by Zhang\etal~\cite{Zhang2020} in a prior benchmark: the SO~\cite{Lin2018} and API~\cite{Gias2021}  data with 4522 resp. 1500 sentences from Stack Overflow posts and the GH data~\cite{Novielli2020} with 7122 sentences from GitHub pull requests and commit messages. All three data sets have three labels, i.e., positive, neutral, and negative sentiment. For each of these data sets, we use the same experimental setting as for the second fine-tuning task, i.e., we conduct 10x10 cross-validation. Since Zhang\etal~\cite{Zhang2020} also the \FSCORE{}, we follow this approach and report the macro averages of the \FSCORE{}, \RECALL{}, and \PRECISION{}, same as for the other use cases. 

\section{Experiments}
\label{sec:results}

We conducted an experiment to evaluate the validity of the SE models seBERT and BERToverflow in comparison to the general domain BERT models. Since seBERT is based on BERT\textsubscript{LARGE} and BERT\textsubscript{BASE}, we use both of these models within our comparison. We follow the procedure outlined in Section~\ref{sec:validation}. This means we first compare the vocabularies. Then, we proceed to look at the validity of the models without fine-tuning through their ability to infer the correct meaning of terms. Finally, we evaluated how the different models perform when fine-tuned with a limited amount of labeled data for a prediction task.

All results, as well as the required scripts and link to the data for pre-training seBERT are available as part of our replication kit.\footnote{\url{https://github.com/smartshark/seBERT}} Additionally, we prepared a playground, which can be used to fill in masked words using the models.\footnote{\url{https://smartshark2.informatik.uni-goettingen.de/sebert/index.html}}

We restrict our comparison to BERT, as this is the most similar general domain model and, therefore, most suitable to explore the need for SE-specific models for SE domain tasks. Please note that we also exclude CodeBERT from this comparison, because CodeBERT is not natural language model, but rather a bi-modal ``natural language - programming language'' model, i.e., for working with source code.\footnote{Nevertheless, because CodeBERT has a natural language part, we included CodeBERT in our playground for filling in masked words, for those interested in the suitability of CodeBERT for language understanding.} Additionally, we use fastText for the comparison with respect to classification tasks, to understand if smaller NLP models may be sufficient.

\subsection{Vocabulary Comparison}
\label{sec:results-vocab}

\begin{table*}
\centering
\begin{tabular}{ll|lll}
\toprule
\multicolumn{2}{c|}{\textbf{BERT}} & & \multicolumn{1}{c}{\textbf{seBERT}} & \multicolumn{1}{c}{\textbf{BERToverflow}}\\
\textbf{Word} & \textbf{Tokenization} & \textbf{Word} & \textbf{Tokenization} &  \textbf{Tokenization} \\
\midrule
bugzilla & bug \#\#zil \#\#la & catholic &  cat \#\#hol \#\#ic & cath \#\#olic \\ 
chromium & ch \#\#rom \#\#ium & drama   &  dram \#\#a  &  drama \\
debug & de \#\#bu \#\#g & infantry &  inf \#\#ant \#\#ry & inf \#\#ant \#\#ry \\ 
jvm & j \#\#v \#\#m & palace &  pal \#\#ace & pal \#\#ace\\ 
refactoring & ref \#\#act \#\#orin \#\#g & woman & wo \#\#man & woman\\  \bottomrule
\end{tabular}
\caption{Example words not included in the BERT or seBERT vocabulary broken down into their WordPieces using the respective tokenizer. \#\# indicates the start of a subword token. }
\label{tab:tokenization}
\end{table*}

The vocabularies of BERT\textsubscript{BASE} and BERT\textsubscript{LARGE} are equal, and we refer to them as BERT vocabulary in the following. The WordPiece overlap between BERT and seBERT is 38.3\%. A major difference between the vocabularies is the number of ``\textbf{\#\#}'' sub-word WordPieces, with seBERT (7430) having almost twice as many as BERT (3285). Most of the words added to the seBERT vocabulary are from the SE domain (e.g. ``bugzilla'', ``jvm'', ``debug'') or internet slang (e.g. ``fanboy''). Lost words are mainly from the geographical (e.g. ``madrid'', ``switzerland'', ``egypt''), religious (e.g. ``jesus'', ``buddha'') or political (e.g. ``president'', ``minister'', ``clinton'') domain.

For words that occur in only one of the two vocabularies, we use the tokenizer of the other model to break them down into their respective WordPieces. We have presented the results for a small subset of words in Table~\ref{tab:tokenization}. As we can see, BERT tokenizes SE domain words inconsistently, e.g. ``bugzilla'' is tokenized into ``bug \#\#zil \#\#la'', but "debug" into ``de \#\#bu \#\#g'', showing that there is no WordPiece for ``\#\#bug''. In addition, in-domain abbreviations such as JVM are unknown and broken down into their individual characters. In contrast, seBERT breaks down general out-of-domain words into in-domain WordPieces, e.g., ``drama'' into dynamic random-access memory (DRAM) or ``infantry'' into ``inf'' and ``ant''. Since the vocabularies are frequency-based, an interesting finding is that ``woman'' is not in the seBERT vocabulary and is tokenized into ``wo \#\#man''. This implies that ``woman'' is used less frequently within the domain, highlighting the ethical issues that can arise in machine learning. The full lists of out-of-vocabulary words and their tokenizations are available as supplemental material in the seBERT replication repository.

Since the BERToverflow vocabulary is cased, a direct comparison of the overlap without further processing is not feasible. Instead, we lower case the vocabulary of BERToverflow and remove duplicates. Since the vocabulary of BERToverflow is larger than that of (se)BERT,\footnote{We write (se)BERT to highlight that these statements are true for both BERT and seBERT, which have vocabularies of the same size.} we still cannot directly compute the overlap, because this is a ratio with respect to the vocabulary size. The larger BERToverflow should contain more words from the other vocabularies (larger ratio), while in turn the smaller the (se)BERT vocabularies should cover a smaller ratio of terms in the BERToverflow vocabulary. Instead, we calculate i) the ratio of WordPieces of the (se)BERT vocabulary within the uncased BERToverflow vocabulary and ii) the ratio of uncased BERToverflow WordPieces within the BERT vocabulary. Thus, consider the direction of the comparison, when considering ratios.

After lower casing, there are 58,854 unique WordPiece tokens. 24,263 of these tokens are also in the seBERT vocabulary. This means seBERT covers about 40\% of the uncased BERToverflow vocabulary and the uncased BERToverflow covers about 80\% of the seBERT vocabulary. That BERToverflow has more tokens is not surprising, due to the larger vocabulary size. While the overlap of 80\% of the seBERT tokens with BERToverflow is substantial, we would have expected an even larger overlap, given that BERToverflow has about twice the amount of uncased tokens. This should be sufficient to more or less cover the seBERT vocabulary, assuming that the textual data from Stack Overflow on which BERToverflow is pre-trained, is representative for the SE domain. A manual check revealed that many of the missing tokens are related to tools, e.g., ``matplot``, ``xenial'' or ``hashicorp''. We can only speculate regarding the reason for this. One possible explanation is that tools are less frequently discussed on Stack Overflow than, e.g., programming languages. However, while this may the case to a certain degree, questions regarding the usage of specific tools are common in Stack Overflow. Another possible explanation is the casing that is used when writing tools. For example, users may write ``HashiCorp'', ``Hashicorp'', or ``hashicorp''. Within the uncased seBERT vocabulary, this would not make a difference and all occurrences would be counted together, possibly leading to larger importance and the inclusion in the vocabulary.

We also manually evaluated which additional terms are within the BERToverflow vocabulary that are not within the seBERT vocabulary. In addition to more terms from the general domain (see the comparison to BERT below), we noticed two additional aspects. First, there were many Unicode tokens, such as special characters only used in non-English texts. Second, there were many terms that seem to come from code snippets, such as ``strftime''. 

The BERT vocabulary has an overlap of 15,926 WordPiece tokens with the uncased BERToverflow vocabulary. This means that about 50\% of the BERT tokens are within the BERToverflow vocabulary, which is slightly more than the overlap of 38\% between seBERT and BERT. However, this increase is plausible, given the overall larger vocabulary size of BERToverflow. This is also visible in the tokenization, e.g., the terms ``women'' and ``drama'' are within BERToverflow vocabulary, meaning that is covers more general domain concepts that seBERT. 

\subsection{Contextual Comparison}
\label{sec:mlm-results}

Next, we use the capability of the models to predict masked words. Tables~\ref{tab:positive}-\ref{tab:neutral} show ten examples of sentences for the positive, negative, and neutral category. 

\begin{table*}
\centering
\begin{tabular}{p{3.4cm} p{1.5cm} ll ll ll ll}
\toprule
 & & \multicolumn{2}{c}{\textbf{BERT\textsubscript{BASE}}} & \multicolumn{2}{c}{\textbf{BERT\textsubscript{LARGE}}} & \multicolumn{2}{c}{\textbf{BERToverflow}} & \multicolumn{2}{c}{\textbf{seBERT}}  \\ 
\textbf{Sentence} & \textbf{Expectation} & \textbf{Prediction} & \textbf{Prob.} & \textbf{Prediction} & \textbf{Prob.} & \textbf{Prediction} & \textbf{Prob.} & \textbf{Prediction} & \textbf{Prob.} \\
\midrule

1) The [MASK] is thrown when an application attempts to use null in a case where an object is required. & NullPointer-Exception & 
\begin{tabular}[t]{@{}l@{}} rule \\ exception \\ coin \\ flag \\ penalty \end{tabular} &
\begin{tabular}[t]{@{}l@{}} 0.2407 \\ 0.0742 \\ 0.0659 \\ 0.0245 \\ 0.0216 \end{tabular} &
\begin{tabular}[t]{@{}l@{}} value \\ exception \\ coin \\ ball \\ flag \end{tabular} &
\begin{tabular}[t]{@{}l@{}} 0.2356 \\ 0.0804 \\ 0.0342 \\ 0.0318 \\ 0.0312 \end{tabular} &
\begin{tabular}[t]{@{}l@{}} exception \\ error \\  Null- \\ Pointer- \\ Exception \\ NPE \\ Illegal- \\ Argument- \\ Exception \end{tabular} &
\begin{tabular}[t]{@{}l@{}} 0.4718 \\ 0.1229 \\ 0.1161 \\ \\ \\ 0.0995 \\ 0.0403 \end{tabular} &
\begin{tabular}[t]{@{}l@{}} exception \\ error \\ null- \\ pointer- \\ exception \\ npe \\ illegal- \\ argument- \\ exception \end{tabular} &
\begin{tabular}[t]{@{}l@{}} 0.5473 \\ 0.2188 \\ 0.1158 \\ \\ \\ 0.0291 \\ 0.0144 \end{tabular} \\ \noalign{\vspace{4pt}}

2) [MASK] is a proprietary issue tracking product developed by Atlassian that allows bug tracking and agile project management. & Jira & 
\begin{tabular}[t]{@{}l@{}} it \\ agile \\ eclipse \\ flex \\ snap \end{tabular} &
\begin{tabular}[t]{@{}l@{}} 0.0382 \\ 0.0104 \\ 0.0041 \\ 0.0035 \\ 0.0031 \end{tabular} &
\begin{tabular}[t]{@{}l@{}}it \\ bug \\ eclipse \\ echo \\ spark \end{tabular} &
\begin{tabular}[t]{@{}l@{}} 0.0148 \\ 0.0074 \\ 0.0052 \\ 0.0046 \\ 0.0035 \end{tabular} &
\begin{tabular}[t]{@{}l@{}} Jira \\ Redmine \\ JIRA \\ It \\ There \end{tabular} &
\begin{tabular}[t]{@{}l@{}} 0.4914 \\ 0.1834 \\ 0.099 \\ 0.0398 \\ 0.0281 \end{tabular} &
\begin{tabular}[t]{@{}l@{}} jira \\ there \\ zenhub \\ it \\ bugzilla \end{tabular} &
\begin{tabular}[t]{@{}l@{}} 0.4954 \\ 0.0837 \\ 0.0763 \\ 0.0488 \\ 0.0463 \end{tabular} \\ \noalign{\vspace{4pt}}

3) [MASK] is a software tool for automating software build processes. & build automation tools &
\begin{tabular}[t]{@{}l@{}} it \\ agile \\ this \\ build \\ gem \end{tabular} &
\begin{tabular}[t]{@{}l@{}} 0.1437 \\ 0.0071 \\ 0.0062 \\ 0.0055 \\ 0.0055 \end{tabular} &
\begin{tabular}[t]{@{}l@{}} it \\ build \\ eclipse \\ builder \\ agile \end{tabular} &
\begin{tabular}[t]{@{}l@{}} 0.0396 \\ 0.0294 \\ 0.0107 \\ 0.0075 \\ 0.0067 \end{tabular} &
\begin{tabular}[t]{@{}l@{}} CMake \\ make \\ Make \\ Ant \\ Maven \end{tabular} &
\begin{tabular}[t]{@{}l@{}} 0.1541 \\ 0.0898 \\ 0.0644 \\ 0.0595 \\ 0.0558 \end{tabular} &
\begin{tabular}[t]{@{}l@{}} jenkins \\ make \\ ninja \\ it \\ ant \end{tabular} &
\begin{tabular}[t]{@{}l@{}} 0.1907 \\ 0.0891 \\ 0.0676 \\ 0.058 \\ 0.0536 \end{tabular} \\ \noalign{\vspace{4pt}}

4) Pathlib is a python library used for handeling [MASK]. & paths &
\begin{tabular}[t]{@{}l@{}} applications \\ systems \\ programs \\ software \\ data \end{tabular} &
\begin{tabular}[t]{@{}l@{}} 0.0699 \\ 0.0431 \\ 0.0359 \\ 0.0351 \\ 0.0296 \end{tabular} &
\begin{tabular}[t]{@{}l@{}} applications \\ software \\ programs \\ languages \\ systems \end{tabular} &
\begin{tabular}[t]{@{}l@{}} 0.1509 \\ 0.1505 \\ 0.078 \\ 0.0631 \\ 0.052 \end{tabular} &
\begin{tabular}[t]{@{}l@{}} paths \\ files \\ urls \\ URLs \\ pathnames \end{tabular} &
\begin{tabular}[t]{@{}l@{}} 0.429 \\ 0.0789 \\ 0.0664 \\ 0.0391 \\ 0.0384 \end{tabular} &
\begin{tabular}[t]{@{}l@{}} paths \\ path \\ directories \\ filenames \\ strings \end{tabular} &
\begin{tabular}[t]{@{}l@{}} 0.8727 \\ 0.0338 \\ 0.0237 \\ 0.0168 \\ 0.0088 \end{tabular} \\ \noalign{\vspace{4pt}}

5) The solution posted by [USER] is [MASK] helpful. :) & positive adverb &
\begin{tabular}[t]{@{}l@{}} very \\ extremely \\ quite \\ always \\ more \end{tabular} &
\begin{tabular}[t]{@{}l@{}} 0.5632 \\ 0.0656 \\ 0.0424 \\ 0.0313 \\ 0.0273 \end{tabular} &
\begin{tabular}[t]{@{}l@{}} very \\ not \\ always \\ also \\ most \end{tabular} &
\begin{tabular}[t]{@{}l@{}} 0.3407 \\ 0.1755 \\ 0.0624 \\ 0.0593 \\ 0.0429 \end{tabular} &
\begin{tabular}[t]{@{}l@{}} very \\ really \\ also \\ more \\ quite \end{tabular} &
\begin{tabular}[t]{@{}l@{}} 0.5488 \\ 0.1374 \\ 0.0734 \\ 0.027 \\ 0.0237 \end{tabular} &
\begin{tabular}[t]{@{}l@{}} very \\ really \\ quite \\ also \\ super \end{tabular} &
\begin{tabular}[t]{@{}l@{}} 0.5719 \\ 0.1533 \\ 0.0729 \\ 0.0458 \\ 0.0166 \end{tabular} \\ \noalign{\vspace{4pt}}

6) The solution posted by [USER] is [MASK] helpful. :( & negative adverb &
\begin{tabular}[t]{@{}l@{}} very \\ extremely \\ quite \\ also \\ always \end{tabular} &
\begin{tabular}[t]{@{}l@{}} 0.5254 \\ 0.0659 \\ 0.0418 \\ 0.0349 \\ 0.0338 \end{tabular} &
\begin{tabular}[t]{@{}l@{}} very \\ not \\ also \\ always \\ most \end{tabular} &
\begin{tabular}[t]{@{}l@{}} 0.2764 \\ 0.1767 \\ 0.0829 \\ 0.0782 \\ 0.0455 \end{tabular} &
\begin{tabular}[t]{@{}l@{}} not \\ very \\ really \\ never \\ also \end{tabular} &
\begin{tabular}[t]{@{}l@{}} 0.9182 \\ 0.0331 \\ 0.0087 \\ 0.0031 \\ 0.0031 \end{tabular} &
\begin{tabular}[t]{@{}l@{}} not \\ no \\ very \\ really \\ never \end{tabular} &
\begin{tabular}[t]{@{}l@{}} 0.9761 \\ 0.0075 \\ 0.0038 \\ 0.0011 \\ 0.0009 \end{tabular} \\ \noalign{\vspace{4pt}}

7) [MASK], is a provider of Internet hosting for software development and version control using Git. & github, gitlab &
\begin{tabular}[t]{@{}l@{}} microsoft \\ net \\ oracle \\ org \\ apache \end{tabular} &
\begin{tabular}[t]{@{}l@{}} 0.0158 \\ 0.0136 \\ 0.0134 \\ 0.0127 \\ 0.0122 \end{tabular} &
\begin{tabular}[t]{@{}l@{}} net \\ apache \\ parallels \\ foundry \\ radius \end{tabular} &
\begin{tabular}[t]{@{}l@{}} 0.0127 \\ 0.011 \\ 0.0097 \\ 0.0073 \\ 0.0061 \end{tabular} &
\begin{tabular}[t]{@{}l@{}} Github \\ GitHub \\ Bitbucket \\ BitBucket \\ Assembla \end{tabular} &
\begin{tabular}[t]{@{}l@{}} 0.222 \\ 0.2186 \\ 0.0568 \\ 0.0509 \\ 0.0413 \end{tabular} &
\begin{tabular}[t]{@{}l@{}} github \\ gitlab \\ git \\ sourceforge \\ bitbucket \end{tabular} &
\begin{tabular}[t]{@{}l@{}} 0.3096 \\ 0.0397 \\ 0.0382 \\ 0.0309 \\ 0.0302 \end{tabular} \\ \noalign{\vspace{4pt}}

8) In object-oriented programming, a [MASK] is an extensible program-code-template for creating objects. & class &
\begin{tabular}[t]{@{}l@{}} template \\ class \\ object \\ model \\ gui \end{tabular} &
\begin{tabular}[t]{@{}l@{}} 0.1027 \\ 0.0391 \\ 0.0356 \\ 0.0219 \\ 0.0145 \end{tabular} &
\begin{tabular}[t]{@{}l@{}} template \\ prototype \\ module \\ class \\ construct \end{tabular} &
\begin{tabular}[t]{@{}l@{}} 0.6763 \\ 0.0533 \\ 0.0258 \\ 0.0136 \\ 0.0106 \end{tabular} &
\begin{tabular}[t]{@{}l@{}} constructor \\ class \\ factory \\ prototype \\ Factory \end{tabular} &
\begin{tabular}[t]{@{}l@{}} 0.3849 \\ 0.2664 \\ 0.1435 \\ 0.038 \\ 0.0182 \end{tabular} &
\begin{tabular}[t]{@{}l@{}} class \\ constructor \\ factory \\ metaclass \\ template \end{tabular} &
\begin{tabular}[t]{@{}l@{}} 0.4991 \\ 0.221 \\ 0.0669 \\ 0.0487 \\ 0.0316 \end{tabular} \\ \noalign{\vspace{4pt}}

9) I have to discuss this with the other [MASK]. & developers & 
\begin{tabular}[t]{@{}l@{}} elders \\ girls \\ men \\ officers \\ council \end{tabular} &
\begin{tabular}[t]{@{}l@{}} 0.0703 \\ 0.0677 \\ 0.0622 \\ 0.0526 \\ 0.0453 \end{tabular} &
\begin{tabular}[t]{@{}l@{}} men \\ members \\ elders \\ officers \\ people \end{tabular} &
\begin{tabular}[t]{@{}l@{}} 0.0913 \\ 0.0407 \\ 0.0399 \\ 0.0265 \\ 0.024 \end{tabular} &
\begin{tabular}[t]{@{}l@{}} guys \\ people \\ developers \\ person \\ users \end{tabular} &
\begin{tabular}[t]{@{}l@{}} 0.1364 \\ 0.1074 \\ 0.0918 \\ 0.0533 \\ 0.0497 \end{tabular} &
\begin{tabular}[t]{@{}l@{}} developers \\ team \\ maintainers \\ people \\ devs \end{tabular} &
\begin{tabular}[t]{@{}l@{}} 0.1159 \\ 0.1114 \\ 0.0797 \\ 0.0796 \\ 0.0744 \end{tabular} \\ \noalign{\vspace{4pt}}

10) This is a [MASK] bug, please address it asap. & critical, major & 
\begin{tabular}[t]{@{}l@{}} serious \\ new \\ major \\ persistent \\ big \end{tabular} &
\begin{tabular}[t]{@{}l@{}} 0.286 \\ 0.1166 \\ 0.0699 \\ 0.0473 \\ 0.031 \end{tabular} &
\begin{tabular}[t]{@{}l@{}} security \\ serious \\ minor \\ major \\ nasty \end{tabular} &
\begin{tabular}[t]{@{}l@{}} 0.0956 \\ 0.0807 \\ 0.0458 \\ 0.0397 \\ 0.0298 \end{tabular} &
\begin{tabular}[t]{@{}l@{}} known \\ know \\ chrome \\ browser \\ common \end{tabular} &
\begin{tabular}[t]{@{}l@{}} 0.5999 \\ 0.029 \\ 0.02 \\ 0.014 \\ 0.0118 \end{tabular} &
\begin{tabular}[t]{@{}l@{}} critical \\ serious \\ major \\ big \\ real \end{tabular} &
\begin{tabular}[t]{@{}l@{}} 0.6915 \\ 0.1491 \\ 0.0412 \\ 0.0270 \\ 0.0099 \end{tabular} \\ \noalign{\vspace{2pt}}

\bottomrule
\end{tabular}
\caption{Prediction of words for [MASK] tokens. Results for the \textit{positive} category, i.e., sentences where we expect that BERToverflow and seBERT perform better than BERT.}
\label{tab:positive}
\end{table*}

\begin{table*}
\centering
\begin{tabular}{p{3.1cm} p{1.5cm} ll ll ll ll}
\toprule
 & & \multicolumn{2}{c}{\textbf{BERT\textsubscript{BASE}}} & \multicolumn{2}{c}{\textbf{BERT\textsubscript{LARGE}}} & \multicolumn{2}{c}{\textbf{BERToverflow}} & \multicolumn{2}{c}{\textbf{seBERT}}  \\ 
\textbf{Sentence} & \textbf{Expectation} & \textbf{Prediction} & \textbf{Prob.} & \textbf{Prediction} & \textbf{Prob.} & \textbf{Prediction} & \textbf{Prob.} & \textbf{Prediction} & \textbf{Prob.} \\
\midrule

11) A [MASK] crawled across her leg, and she swiped it away. & bug &
\begin{tabular}[t]{@{}l@{}} spider \\ tear \\ hand \\ bug \\ mosquito \end{tabular} &
\begin{tabular}[t]{@{}l@{}} 0.1417 \\ 0.1277 \\ 0.0841 \\ 0.0834 \\ 0.0534 \end{tabular} &
\begin{tabular}[t]{@{}l@{}} bug \\ spider \\ tear \\ flea \\ fly \end{tabular} &
\begin{tabular}[t]{@{}l@{}} 0.2087 \\ 0.1676 \\ 0.1463 \\ 0.048 \\ 0.0447 \end{tabular} &
\begin{tabular}[t]{@{}l@{}} person \\ car \\ dog \\ bird \\ fox \end{tabular} &
\begin{tabular}[t]{@{}l@{}} 0.1068 \\ 0.0561 \\ 0.0559 \\ 0.0405 \\ 0.0397 \end{tabular} &
\begin{tabular}[t]{@{}l@{}} man \\ person \\ friend \\ girl \\ monkey \end{tabular} &
\begin{tabular}[t]{@{}l@{}} 0.1042 \\ 0.0518 \\ 0.0472 \\ 0.0299 \\ 0.0248 \end{tabular} \\ \noalign{\vspace{4pt}}

12) Can you open the [MASK], please? It's hot in here. & window, door &
\begin{tabular}[t]{@{}l@{}} door \\ window \\ windows \\ blinds \\ curtains \end{tabular} &
\begin{tabular}[t]{@{}l@{}} 0.7435 \\ 0.1227 \\ 0.0349 \\ 0.0171 \\ 0.0113 \end{tabular} &
\begin{tabular}[t]{@{}l@{}} door \\ window \\ fridge \\ doors \\ gate \end{tabular} &
\begin{tabular}[t]{@{}l@{}} 0.8776 \\ 0.0619 \\ 0.0058 \\ 0.0057 \\ 0.0055 \end{tabular} &
\begin{tabular}[t]{@{}l@{}} link \\ file \\ site \\ page \\ url \end{tabular} &
\begin{tabular}[t]{@{}l@{}} 0.3264 \\ 0.0699 \\ 0.0411 \\ 0.0395 \\ 0.0307 \end{tabular} &
\begin{tabular}[t]{@{}l@{}} pr \\ issue \\ door \\ file \\ link \end{tabular} &
\begin{tabular}[t]{@{}l@{}} 0.2829 \\ 0.1386 \\ 0.0576 \\ 0.0466 \\ 0.0352 \end{tabular} \\ \noalign{\vspace{4pt}}

13) The reticulated python is among the few [MASK] that prey on humans. & snakes & 
\begin{tabular}[t]{@{}l@{}} snakes \\ species \\ reptiles \\ animals \\ lizards \end{tabular} &
\begin{tabular}[t]{@{}l@{}} 0.6433 \\ 0.1645 \\ 0.1055 \\ 0.0278 \\ 0.0269 \end{tabular} &
\begin{tabular}[t]{@{}l@{}} snakes \\ reptiles \\ animals \\ species \\ mammals \end{tabular} &
\begin{tabular}[t]{@{}l@{}} 0.803 \\ 0.0779 \\ 0.0696 \\ 0.0223 \\ 0.0075 \end{tabular} &
\begin{tabular}[t]{@{}l@{}} languages \\ things \\ tools \\ people \\ programmers \end{tabular} & 
\begin{tabular}[t]{@{}l@{}} 0.7376 \\ 0.0836 \\ 0.0087 \\ 0.0084 \\ 0.0073 \end{tabular} &
\begin{tabular}[t]{@{}l@{}} things \\ languages \\ tools \\ packages \\ programs \end{tabular} &
\begin{tabular}[t]{@{}l@{}} 0.5227 \\ 0.1625 \\ 0.0632 \\ 0.0348 \\ 0.0255 \end{tabular} \\ \noalign{\vspace{4pt}}

14) ``I have a [MASK] request for you.'' He said to the waiter. & special &
\begin{tabular}[t]{@{}l@{}} special \\ business \\ personal \\ new \\ specific \end{tabular} &
\begin{tabular}[t]{@{}l@{}} 0.4258 \\ 0.0859 \\ 0.0809 \\ 0.0318 \\ 0.0225 \end{tabular} &
\begin{tabular}[t]{@{}l@{}} special \\ specific \\ new \\ small \\ simple \end{tabular} &
\begin{tabular}[t]{@{}l@{}} 0.8448 \\ 0.0198 \\ 0.017 \\ 0.0166 \\ 0.0128 \end{tabular} &
\begin{tabular}[t]{@{}l@{}} new \\ special \\ test \\ friend \\ support \end{tabular} &
\begin{tabular}[t]{@{}l@{}} 0.115 \\ 0.0361 \\ 0.0257 \\ 0.0224 \\ 0.0213 \end{tabular} &
\begin{tabular}[t]{@{}l@{}} pull \\ feature \\ change \\ similar \\ merge \end{tabular} &
\begin{tabular}[t]{@{}l@{}} 0.7016 \\ 0.0432 \\ 0.0171 \\ 0.0146 \\ 0.0139 \end{tabular} \\ \noalign{\vspace{4pt}}

15) He was admitting to a [MASK] he didn't commit, knowing it was somebody else who did it. & crime &
\begin{tabular}[t]{@{}l@{}} crime \\ murder \\ sin \\ lie \\ suicide \end{tabular} &
\begin{tabular}[t]{@{}l@{}} 0.7861 \\ 0.1162 \\ 0.0484 \\ 0.0084 \\ 0.0056 \end{tabular} &
\begin{tabular}[t]{@{}l@{}} crime \\ murder \\ sin \\ felony \\ lie \end{tabular} &
\begin{tabular}[t]{@{}l@{}} 0.9565 \\ 0.0358 \\ 0.005 \\ 0.0008 \\ 0.0005 \end{tabular} &
\begin{tabular}[t]{@{}l@{}} commit \\ change \\ mistake \\ fact \\ file \end{tabular} &
\begin{tabular}[t]{@{}l@{}} 0.1497 \\ 0.1223 \\ 0.0844 \\ 0.0629 \\ 0.044 \end{tabular} &
\begin{tabular}[t]{@{}l@{}} change \\ commit \\ fix \\ file \\ code \end{tabular} &
\begin{tabular}[t]{@{}l@{}} 0.2475 \\ 0.0974 \\ 0.0833 \\ 0.0464 \\ 0.0372 \end{tabular} \\ \noalign{\vspace{4pt}}

16) It's an incurable, terminal [MASK]. & disease & 
\begin{tabular}[t]{@{}l@{}} disease \\ condition \\ illness \\ death \\ cancer \end{tabular} &
\begin{tabular}[t]{@{}l@{}} 0.4208 \\ 0.3208 \\ 0.125 \\ 0.0201 \\ 0.0193 \end{tabular} &
\begin{tabular}[t]{@{}l@{}} disease \\ illness \\ condition \\ cancer \\ disorder \end{tabular} &
\begin{tabular}[t]{@{}l@{}} 0.8134 \\ 0.0879 \\ 0.0579 \\ 0.0203 \\ 0.0049 \end{tabular} &
\begin{tabular}[t]{@{}l@{}} error \\ operation \\ command \\ problem \\ ) \end{tabular} &
\begin{tabular}[t]{@{}l@{}} 0.0777 \\ 0.0639 \\ 0.0517 \\ 0.0336 \\ 0.0288 \end{tabular} &
\begin{tabular}[t]{@{}l@{}} issue \\ problem \\ bug \\ character \\ effect \end{tabular} &
\begin{tabular}[t]{@{}l@{}} 0.0605 \\ 0.0523 \\ 0.0443 \\ 0.0377 \\ 0.0273 \end{tabular} \\ \noalign{\vspace{4pt}}

17) The dentist said I need to have a root [MASK]. & canal &
\begin{tabular}[t]{@{}l@{}} canal \\ beer \\ problem \\ out \\ cellar \end{tabular} &
\begin{tabular}[t]{@{}l@{}} 0.8847 \\ 0.0307 \\ 0.0167 \\ 0.005 \\ 0.003 \end{tabular} &
\begin{tabular}[t]{@{}l@{}} canal \\ beer \\ stop \\ break \\ problem \end{tabular} &
\begin{tabular}[t]{@{}l@{}} 0.9827 \\ 0.01 \\ 0.0009 \\ 0.0005 \\ 0.0005 \end{tabular} &
\begin{tabular}[t]{@{}l@{}} account \\ certificate \\ user \\ node \\ access \end{tabular} &
\begin{tabular}[t]{@{}l@{}} 0.1919 \\ 0.0959 \\ 0.0825 \\ 0.0588 \\ 0.0474 \end{tabular} &
\begin{tabular}[t]{@{}l@{}} password \\ user \\ account \\ access \\ certificate \end{tabular} &
\begin{tabular}[t]{@{}l@{}} 0.1476 \\ 0.1431 \\ 0.089 \\ 0.0522 \\ 0.0475 \end{tabular} \\ \noalign{\vspace{4pt}}

18) Everything was covered with a fine layer of [MASK]. & dust, snow &
\begin{tabular}[t]{@{}l@{}} dust \\ dirt \\ snow \\ paint \\ powder \end{tabular} &
\begin{tabular}[t]{@{}l@{}} 0.4895 \\ 0.1035 \\ 0.0625 \\ 0.0432 \\ 0.01 \end{tabular} &
\begin{tabular}[t]{@{}l@{}} dust \\ snow \\ dirt \\ ice \\ sand \end{tabular} &
\begin{tabular}[t]{@{}l@{}} 0.7656 \\ 0.0813 \\ 0.0283 \\ 0.0199 \\ 0.0122 \end{tabular} &
\begin{tabular}[t]{@{}l@{}} transparency \\ abstraction \\ code \\ complexity \\ confidence \end{tabular} &
\begin{tabular}[t]{@{}l@{}} 0.0518 \\ 0.0417 \\ 0.0408 \\ 0.0323 \\ 0.032 \end{tabular} &
\begin{tabular}[t]{@{}l@{}} abstraction \\ coverage \\ testing \\ detail \\ documentation \end{tabular} &
\begin{tabular}[t]{@{}l@{}} 0.1231 \\ 0.095 \\ 0.0894 \\ 0.0671 \\ 0.0636 \end{tabular} \\ \noalign{\vspace{4pt}}

19) There was not a single cloud in the [MASK]. & sky & 
\begin{tabular}[t]{@{}l@{}} sky \\ air \\ distance \\ skies \\ room \end{tabular} &
\begin{tabular}[t]{@{}l@{}} 0.9009 \\ 0.0748 \\ 0.003 \\ 0.0021 \\ 0.0017 \\ 0.0032 \end{tabular} &
\begin{tabular}[t]{@{}l@{}} sky \\ heavens \\ distance \\ air \\ east \end{tabular} &
\begin{tabular}[t]{@{}l@{}} 0.9425 \\ 0.0104 \\ 0.0076 \\ 0.0041 \\ 0.0032 \end{tabular} &
\begin{tabular}[t]{@{}l@{}} list \\ cloud \\ world \\ center \\ database \end{tabular} &
\begin{tabular}[t]{@{}l@{}} 0.0797 \\ 0.0361 \\ 0.0333 \\ 0.027 \\ 0.0263 \end{tabular} &
\begin{tabular}[t]{@{}l@{}} database \\ list \\ dataset \\ file \\ cloud \end{tabular} &
\begin{tabular}[t]{@{}l@{}} 0.0686 \\ 0.0663 \\ 0.0391 \\ 0.0358 \\ 0.0357 \end{tabular} \\ \noalign{\vspace{4pt}}

20) What does it say in your [MASK] cookie? & fortune &
\begin{tabular}[t]{@{}l@{}} fortune \\ chocolate \\ little \\ next \\ sugar \end{tabular} &
\begin{tabular}[t]{@{}l@{}} 0.4846 \\ 0.0534 \\ 0.0289 \\ 0.0207 \\ 0.0206 \end{tabular} & 
\begin{tabular}[t]{@{}l@{}} fortune \\ favorite \\ next \\ chocolate \\ sugar \end{tabular} &
\begin{tabular}[t]{@{}l@{}} 0.295 \\ 0.2597 \\ 0.0484 \\ 0.0284 \\ 0.0261 \end{tabular} & 
\begin{tabular}[t]{@{}l@{}} browser \\ session \\ firebug \\ firefox \\ debug \end{tabular} & 
\begin{tabular}[t]{@{}l@{}} 0.1262 \\ 0.083 \\ 0.07 \\ 0.0508 \\ 0.032 \end{tabular} & 
\begin{tabular}[t]{@{}l@{}} session \\ browser \\ auth \\ cookie \\ login \end{tabular} & 
\begin{tabular}[t]{@{}l@{}} 0.4255 \\ 0.0994 \\ 0.0572 \\ 0.0318 \\ 0.0195 \end{tabular} \\ \noalign{\vspace{4pt}}

\bottomrule
\end{tabular}
\caption{Prediction of words for [MASK] tokens. Results for the \textit{negative} category, i.e., sentences where we expect that BERToverflow and seBERT perform worse than BERT.}
\label{tab:negative}
\end{table*}

\begin{table*}
\centering
\begin{tabular}{p{3.0cm} p{1.49cm} ll ll ll ll}
\toprule
 & & \multicolumn{2}{c}{\textbf{BERT\textsubscript{BASE}}} & \multicolumn{2}{c}{\textbf{BERT\textsubscript{LARGE}}} & \multicolumn{2}{c}{\textbf{BERToverflow}} & \multicolumn{2}{c}{\textbf{seBERT}}  \\ 
\textbf{Sentence} & \textbf{Expectation} & \textbf{Prediction} & \textbf{Prob.} & \textbf{Prediction} & \textbf{Prob.} & \textbf{Prediction} & \textbf{Prob.} & \textbf{Prediction} & \textbf{Prob.} \\
\midrule

21) We can [MASK] in person if you have any specific questions. & meet &
\begin{tabular}[t]{@{}l@{}} talk \\ speak \\ meet \\ chat \\ visit \end{tabular} &
\begin{tabular}[t]{@{}l@{}} 0.4357 \\ 0.207 \\ 0.0971 \\ 0.0746 \\ 0.0266 \end{tabular} &
\begin{tabular}[t]{@{}l@{}} meet \\ speak \\ talk \\ discuss \\ communicate \end{tabular} &
\begin{tabular}[t]{@{}l@{}} 0.695 \\ 0.1173 \\ 0.1026 \\ 0.0208 \\ 0.0057 \end{tabular} &
\begin{tabular}[t]{@{}l@{}} help \\ be \\ edit \\ assist \\ answer \end{tabular} &
\begin{tabular}[t]{@{}l@{}} 0.2649 \\ 0.066 \\ 0.0566 \\ 0.0553 \\ 0.0477 \end{tabular} &
\begin{tabular}[t]{@{}l@{}} chat \\ talk \\ discuss \\ meet \\ speak \end{tabular} &
\begin{tabular}[t]{@{}l@{}} 0.5492 \\ 0.1941 \\ 0.1025 \\ 0.0913 \\ 0.0064 \end{tabular} \\ \noalign{\vspace{4pt}}

22) [MASK] its name, vitamin D is not a vitamin. Instead, it is a hormone that promotes the absorption of calcium in the body. & despite &
\begin{tabular}[t]{@{}p{1.5cm}@{}} despite \\ whatever \\ unlike \\ in \\ notwith- standing \end{tabular} &
\begin{tabular}[t]{@{}l@{}} 0.999 \\ 0.0002 \\ 0.0002 \\ 0.0001 \\ 0.0001 \end{tabular} &
\begin{tabular}[t]{@{}l@{}} despite \\ unlike \\ notwithstanding \\ like \\ whatever \end{tabular} &
\begin{tabular}[t]{@{}l@{}} 0.998 \\ 0.0016 \\ 0.0002 \\ 0.0001 \\ 0.0 \end{tabular} &
\begin{tabular}[t]{@{}l@{}} Despite \\ despite \\ by \\ By \\ In \end{tabular} &
\begin{tabular}[t]{@{}l@{}} 0.7482 \\ 0.1184 \\ 0.0387 \\ 0.0212 \\ 0.01 \end{tabular} &
\begin{tabular}[t]{@{}l@{}} despite \\ in \\ from \\ unlike \\ by \end{tabular} &
\begin{tabular}[t]{@{}l@{}} 0.9396 \\ 0.0169 \\ 0.0099 \\ 0.0095 \\ 0.0068 \end{tabular} \\ \noalign{\vspace{4pt}}

23) Would all those in favour please raise their [MASK]? & hands & 
\begin{tabular}[t]{@{}l@{}} hands \\ hand \\ voices \\ arms \\ fists \end{tabular} &
\begin{tabular}[t]{@{}l@{}} 0.5325 \\ 0.122 \\ 0.1177 \\ 0.031 \\ 0.0124 \end{tabular} &
\begin{tabular}[t]{@{}l@{}} hands \\ hand \\ voices \\ voice \\ arms \end{tabular} &
\begin{tabular}[t]{@{}l@{}} 0.537 \\ 0.1543 \\ 0.128 \\ 0.11 \\ 0.0121 \end{tabular} &
\begin{tabular}[t]{@{}l@{}} opinion \\ opinions \\ concerns \\ points \\ views \end{tabular} &
\begin{tabular}[t]{@{}l@{}} 0.2422 \\ 0.1803 \\ 0.0982 \\ 0.0704 \\ 0.0444 \end{tabular} &
\begin{tabular}[t]{@{}l@{}} opinion \\ concerns \\ priority \\ opinions \\ interest \end{tabular} &
\begin{tabular}[t]{@{}l@{}} 0.1235 \\ 0.0798 \\ 0.057 \\ 0.0531 \\ 0.043 \end{tabular} \\ \noalign{\vspace{4pt}}

24) She surprised him with a [MASK]. & something positive &  
\begin{tabular}[t]{@{}l@{}} smile \\ look \\ laugh \\ kiss \\ grin \end{tabular} &
\begin{tabular}[t]{@{}l@{}} 0.4748 \\ 0.1623 \\ 0.0604 \\ 0.0588 \\ 0.0502 \end{tabular} &
\begin{tabular}[t]{@{}l@{}} smile \\ laugh \\ kiss \\ question \\ grin \end{tabular} &
\begin{tabular}[t]{@{}l@{}} 0.4086 \\ 0.2211 \\ 0.105 \\ 0.0673 \\ 0.0369 \end{tabular} &
\begin{tabular}[t]{@{}l@{}} surprise \\ mistake \\ message \\ bug \\ warning \end{tabular} &
\begin{tabular}[t]{@{}l@{}} 0.1815 \\ 0.0606 \\ 0.0551 \\ 0.034 \\ 0.0322 \end{tabular} &
\begin{tabular}[t]{@{}l@{}} bug \\ question \\ comment \\ problem \\ joke \end{tabular} &
\begin{tabular}[t]{@{}l@{}} 0.0604 \\ 0.0485 \\ 0.0197 \\ 0.0191 \\ 0.019 \end{tabular} \\ \noalign{\vspace{4pt}}

25) Whoever is happy will make others [MASK] too. & happy &
\begin{tabular}[t]{@{}l@{}} happy \\ happier \\ , \\ smile \\ sad \end{tabular} &
\begin{tabular}[t]{@{}l@{}} 0.989 \\ 0.0025 \\ 0.0012 \\ 0.0012 \\ 0.0007 \end{tabular} &
\begin{tabular}[t]{@{}l@{}} happy \\ unhappy \\ happier \\ , \\ smile \end{tabular} &
\begin{tabular}[t]{@{}l@{}} 0.9716 \\ 0.0104 \\ 0.0053 \\ 0.0042 \\ 0.002 \end{tabular} &
\begin{tabular}[t]{@{}l@{}} happy \\ unhappy \\ , \\ pleased \\ sad \end{tabular} &
\begin{tabular}[t]{@{}l@{}} 0.9968 \\ 0.0003 \\ 0.0003 \\ 0.0002 \\ 0.0002 \end{tabular} &
\begin{tabular}[t]{@{}l@{}} happy \\ , \\ complain \\ sad \\ comfortable \end{tabular} &
\begin{tabular}[t]{@{}l@{}} 0.9866 \\ 0.0027 \\ 0.0007 \\ 0.0005 \\ 0.0003 \end{tabular} \\ \noalign{\vspace{4pt}}

26) If there are night owls, are there [MASK] owls too? & day &
\begin{tabular}[t]{@{}l@{}} night \\ day \\ other \\ morning \\ bird \end{tabular} &
\begin{tabular}[t]{@{}l@{}} 0.3312 \\ 0.2063 \\ 0.0275 \\ 0.0175 \\ 0.0158 \end{tabular} &
\begin{tabular}[t]{@{}l@{}} day \\ night \\ morning \\ other \\ evening \end{tabular} &
\begin{tabular}[t]{@{}l@{}} 0.5414 \\ 0.2378 \\ 0.0671 \\ 0.0143 \\ 0.0085 \end{tabular} &
\begin{tabular}[t]{@{}l@{}} day \\ night \\ power \\ weather \\ evening \end{tabular} &
\begin{tabular}[t]{@{}l@{}} 0.1331 \\ 0.0527 \\ 0.0352 \\ 0.0178 \\ 0.0149 \end{tabular} &
\begin{tabular}[t]{@{}l@{}} day \\ evening \\ afternoon \\ morning \\ night \end{tabular} &
\begin{tabular}[t]{@{}l@{}} 0.6346 \\ 0.1119 \\ 0.0899 \\ 0.0484 \\ 0.0302 \end{tabular} \\ \noalign{\vspace{4pt}}

27) Never forget, always remember. Always forget, never [MASK]. & remember &
\begin{tabular}[t]{@{}l@{}} forget \\ remember \\ forgot \\ forgotten \\ know \end{tabular} &
\begin{tabular}[t]{@{}l@{}} 0.979 \\ 0.007 \\ 0.0018 \\ 0.0015 \\ 0.0015 \end{tabular} &
\begin{tabular}[t]{@{}l@{}} forget \\ remember \\ forgive \\ know \\ forgot \end{tabular} &
\begin{tabular}[t]{@{}l@{}} 0.9271 \\ 0.0666 \\ 0.0007 \\ 0.0005 \\ 0.0004 \end{tabular} &
\begin{tabular}[t]{@{}l@{}} remember \\ forget \\ trust \\ know \\ read \end{tabular} &
\begin{tabular}[t]{@{}l@{}} 0.4836 \\ 0.3044 \\ 0.0355 \\ 0.0102 \\ 0.0077 \end{tabular} &
\begin{tabular}[t]{@{}l@{}} remember \\ forget \\ recall \\ bother \\ guess \end{tabular} &
\begin{tabular}[t]{@{}l@{}} 0.881 \\ 0.0997 \\ 0.0113 \\ 0.0009 \\ 0.0006 \end{tabular} \\ \noalign{\vspace{4pt}}

28) If you are counting things, start from [MASK]. & 1, one &
\begin{tabular}[t]{@{}l@{}} scratch \\ there \\ bottom \\ one \\ here \end{tabular} &
\begin{tabular}[t]{@{}l@{}} 0.5561 \\ 0.0836 \\ 0.0511 \\ 0.0425 \\ 0.0286 \end{tabular} &
\begin{tabular}[t]{@{}l@{}} one \\ here \\ zero \\ three \\ ten \end{tabular} &
\begin{tabular}[t]{@{}l@{}} 0.1389 \\ 0.1222 \\ 0.0731 \\ 0.064 \\ 0.0497 \end{tabular} &
\begin{tabular}[t]{@{}l@{}} 0 \\ 1 \\ zero \\ 2 \\ there \end{tabular} &
\begin{tabular}[t]{@{}l@{}} 0.3938 \\ 0.293 \\ 0.164 \\ 0.0156 \\ 0.0102 \end{tabular} &
\begin{tabular}[t]{@{}l@{}} zero \\ 0 \\ 1 \\ there \\ 2 \end{tabular} &
\begin{tabular}[t]{@{}l@{}} 0.34 \\ 0.2834 \\ 0.2587 \\ 0.021 \\ 0.0132 \end{tabular} \\ \noalign{\vspace{4pt}}

29) Soccer has really simple rules. It’s not [MASK] science. & rocket &
\begin{tabular}[t]{@{}l@{}} a \\ rocket \\ really \\ just \\ pure \end{tabular} &
\begin{tabular}[t]{@{}l@{}} 0.7412 \\ 0.0334 \\ 0.0227 \\ 0.0225 \\ 0.0147 \end{tabular} &
\begin{tabular}[t]{@{}l@{}} rocket \\ about \\ a \\ even \\ like \end{tabular} &
\begin{tabular}[t]{@{}l@{}} 0.9125 \\ 0.0356 \\ 0.032 \\ 0.0021 \\ 0.0015 \end{tabular} &
\begin{tabular}[t]{@{}l@{}} rocket \\ computer \\ a \\ exact \\ perfect \end{tabular} &
\begin{tabular}[t]{@{}l@{}} 0.8597 \\ 0.0513 \\ 0.0317 \\ 0.0071 \\ 0.0049 \end{tabular} &
\begin{tabular}[t]{@{}l@{}} computer \\ rocket \\ a \\ game \\ about \end{tabular} &
\begin{tabular}[t]{@{}l@{}} 0.4413 \\ 0.3927 \\ 0.0483 \\ 0.0146 \\ 0.0132 \end{tabular} \\ \noalign{\vspace{4pt}}

30) He ran out of [MASK], so he had to stop playing poker. & money, time &
\begin{tabular}[t]{@{}l@{}} money \\ time \\ food \\ funds \\ cash \end{tabular} &
\begin{tabular}[t]{@{}l@{}} 0.8096 \\ 0.0285 \\ 0.0126 \\ 0.0113 \\ 0.0094 \end{tabular} &
\begin{tabular}[t]{@{}l@{}} money \\ cash \\ cards \\ time \\ patience \end{tabular} &
\begin{tabular}[t]{@{}l@{}} 0.7833 \\ 0.0579 \\ 0.0213 \\ 0.0187 \\ 0.0074 \end{tabular} &
\begin{tabular}[t]{@{}l@{}} cards \\ players \\ hands \\ money \\ memory \end{tabular} &
\begin{tabular}[t]{@{}l@{}} 0.4876 \\ 0.0369 \\ 0.0309 \\ 0.0294 \\ 0.0221 \end{tabular} &
\begin{tabular}[t]{@{}l@{}} pokemon \\ memory \\ ram \\ money \\ mana \end{tabular} &
\begin{tabular}[t]{@{}l@{}} 0.1655 \\ 0.0767 \\ 0.0649 \\ 0.0618 \\ 0.0455 \end{tabular} \\ \noalign{\vspace{4pt}}

\bottomrule
\end{tabular}
\caption{Prediction of words for [MASK] tokens. Results for the \textit{neutral} category, i.e., sentences where we expect that BERToverflow and seBERT perform worse than BERT.}
\label{tab:neutral}
\end{table*}

The positive examples in Table~\ref{tab:positive} show that while the general domain BERT models are not as accurate with the identification of words from the within-domain context as the in-domain models BERToverflow and seBERT. This does not mean that BERT outright fails, but rather that the inferred are not completely unrelated, but also not directly on point. In comparison, BERToverflow and seBERT both almost always suggest reasonable completions. However, we also observe that either the larger set of training data or the larger model also allows seBERT to be more accurate in some cases. Our sensitivity analysis with a smaller version of seBERT shows that likely both play a role, as the smaller model is almost equal, but worse in some nuances, i.e., handling of emoticons and opposites. The sensitivity analysis can be found in the supplemental material. The last example gives the strongest indication of the impact of using more data: since BERToverflow was not trained on issue data, but only on Q\&A data from Stack Overflow, seBERT performs better in this context. For this sentence, the general domain results from BERT are actually better than BERToverflow. Our interpretation of this is that this is caused by the context of the BERToverflow data: the criticality of issues is usually not discussed on Stack Overflow. Instead, users may ask regarding a certain problem, with other users responding that the bug is known. Thus, the association of ``This is a [MASK] bug'' with ``known'' makes sense. However, the second part of the sentence further clarifies this context as ``please address this asap''. Such a direct request to the developers does not make sense within the Q\&A context and could, therefore, not be considered by BERToverflow. In comparison, the seBERT data is aware of such direct communication with a development team and correctly infers that the missing word is likely the criticality of the issue, because it should be addressed as soon as possible.

The negative examples in Table~\ref{tab:negative} highlight that models trained with SE data fail, when it comes to understanding pure general domain context. All words are interpreted within the SE context, even if this does not make sense. In comparison, the general domain models were a lot closer to the performance of the SE domain models for the positive examples. When we consider the training corpus, this makes sense: topics like dentists, actual snakes, or the weather are extremely unlikely in our SE corpus, so it is plausible that this does not work at all. On the other hand, Wikipedia also covers software engineering topics. Thus, while this is not the focus of the general domain corpus, it also contains some software engineering data. 

The neutral examples in Table~\ref{tab:neutral} reveal some nuanced differences between all four models. This is the first time that we clearly observe that the BERT\textsubscript{LARGE} model is better a capturing the context than the BERT\textsubscript{BASE} model. In Sentence 28, the smaller models fail to correctly understand the context, while all other models understand this perfectly. The same sentence also reveals a nuanced difference: since we often start to count from zero in computer science, this is also proposed by BERToverflow and seBERT, while BERT\textsubscript{LARGE} says to start at one. We also note that the SE-specific models sometimes misjudge these neutral concepts, e.g, with sentence 23. While the sentence clearly indicates that the context is voting, the term ``raise'' is so strongly associated with opinions within the SE domain that hands are not suggested. This set of sentences also contains the strangest association within our data in Sentence 30: we have no idea why seBERT believes that running out of pokemon is likely. The only explanation we can think of is the tokenization by seBERT of ``poker'' as ``poke\#\#r''. It is possible that the subword ``poke'' has a strong association with ``pokemon'', which leads to this mistake.

\begin{figure*}
\centering
\includegraphics[width=\textwidth]{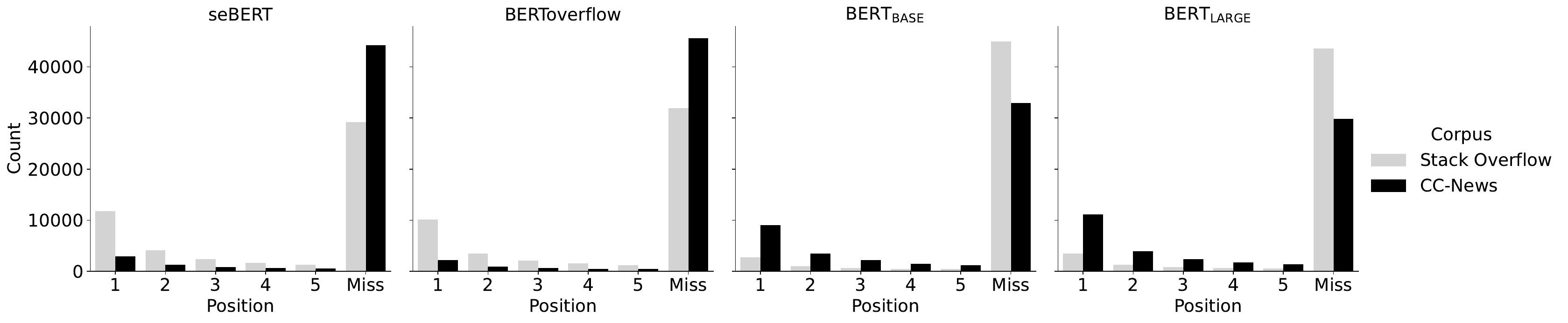}
\caption{Results of the MLM experiment on unknown data. The histograms show the position of the masked word in the top five predicted words of each model, or if the word was missed.}
\label{fig:mlm-results}
\end{figure*}

Figure~\ref{fig:mlm-results} shows the results of our experiments of masking polysemes in the Stack Overflow corpus and CC-News corpus. Overall, we identified 503 words, which appear in at least 100 sentences in each corpus.\footnote{The full list of polysemes can be found in the supplemental material.} Thus, we have $503\cdot 100=50,300$ masked word predictions for each corpus. Overall, the results are in line with our expectation: seBERT and BERToverflow are better for the Stack Overflow corpus, the BERT models are better for CC-News. Moreover, the larger models (seBERT, BERT\textsubscript{LARGE}) perform a bit better than the smaller models (BERToverflow, BERT\textsubscript{BASE}). An interesting aspect of the models is that the results are symmetric: the seBERT results on Stack Overflow are similar to the BERT\textsubscript{LARGE} results on CC-News and the seBERT results on CC-News are similar to the BERT\textsubscript{LARGE} results on Stack Overflow. The same relationship can be observed between BERToverflow and BERT\textsubscript{BASE}. This could mean that the contextual understanding of the SE domain models for the general domain is about as good as that of the general domain models for the SE domain, and vice versa.

\subsection{Prediction Task Comparison}
\begin{figure*}
\centering
\includegraphics[width=0.8\textwidth]{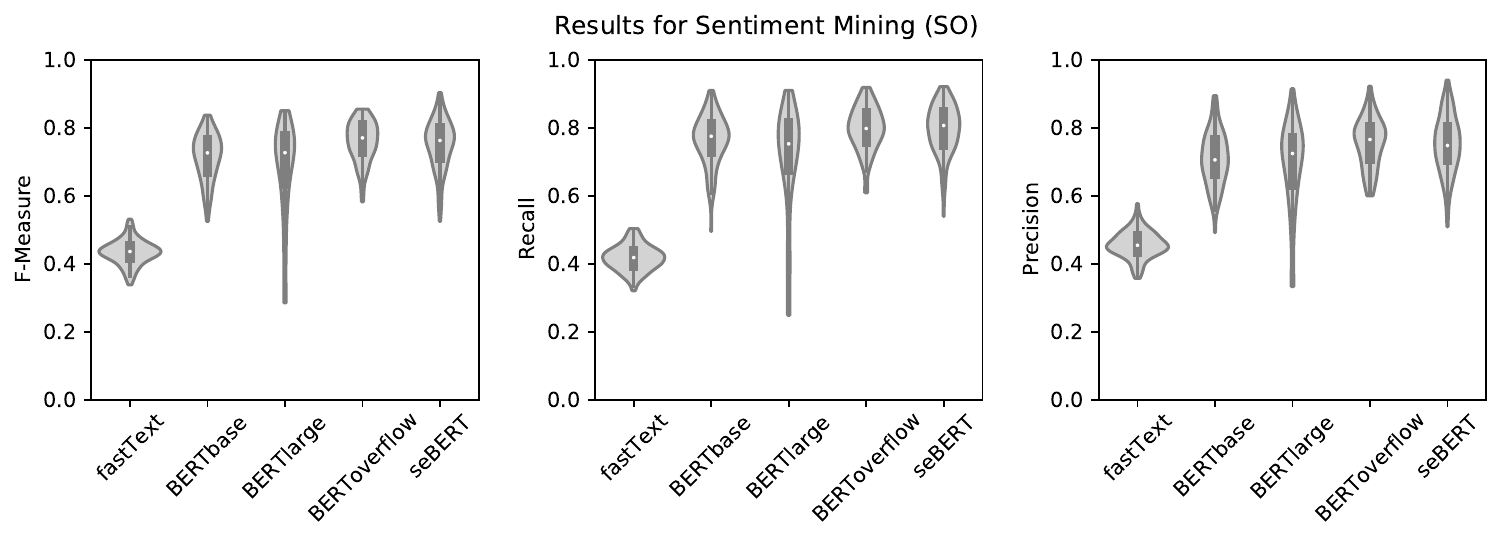}
\caption{Results of the sentiment mining on the SO data.}
\label{fig:results_sentiment_so}
\end{figure*}

\begin{figure*}
\centering
\includegraphics[width=0.8\textwidth]{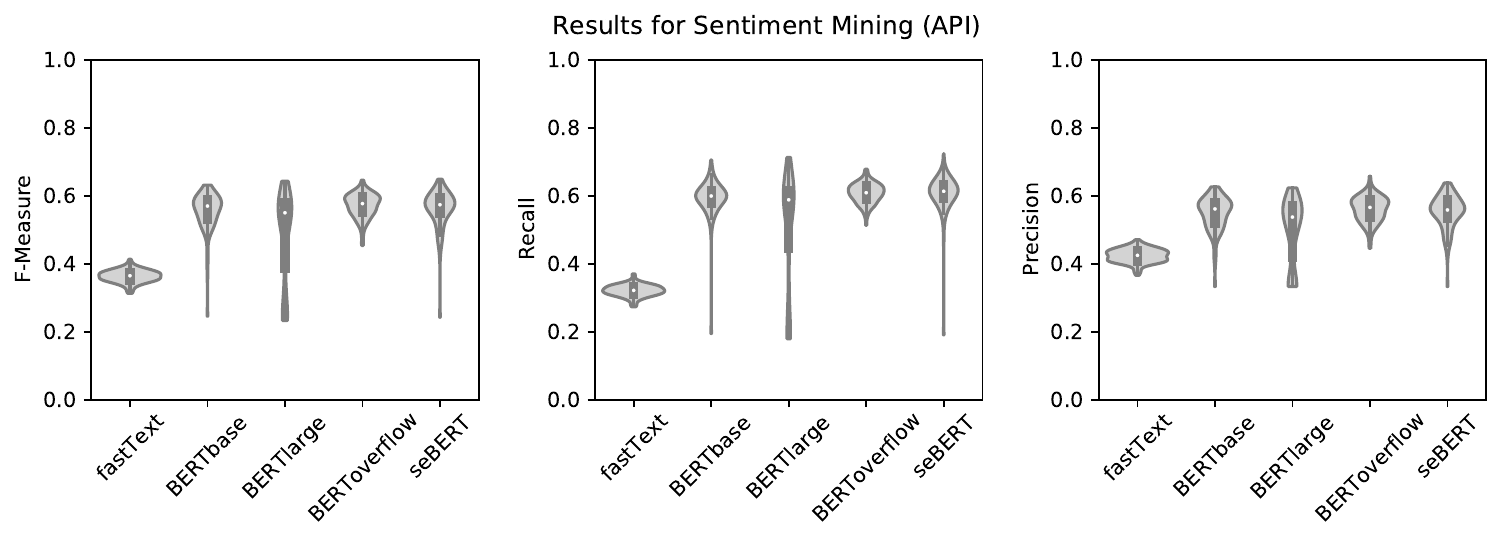}
\caption{Results of the sentiment mining on the API data.}
\label{fig:results_sentiment_api}
\end{figure*}

\begin{figure*}
\centering
\includegraphics[width=0.8\textwidth]{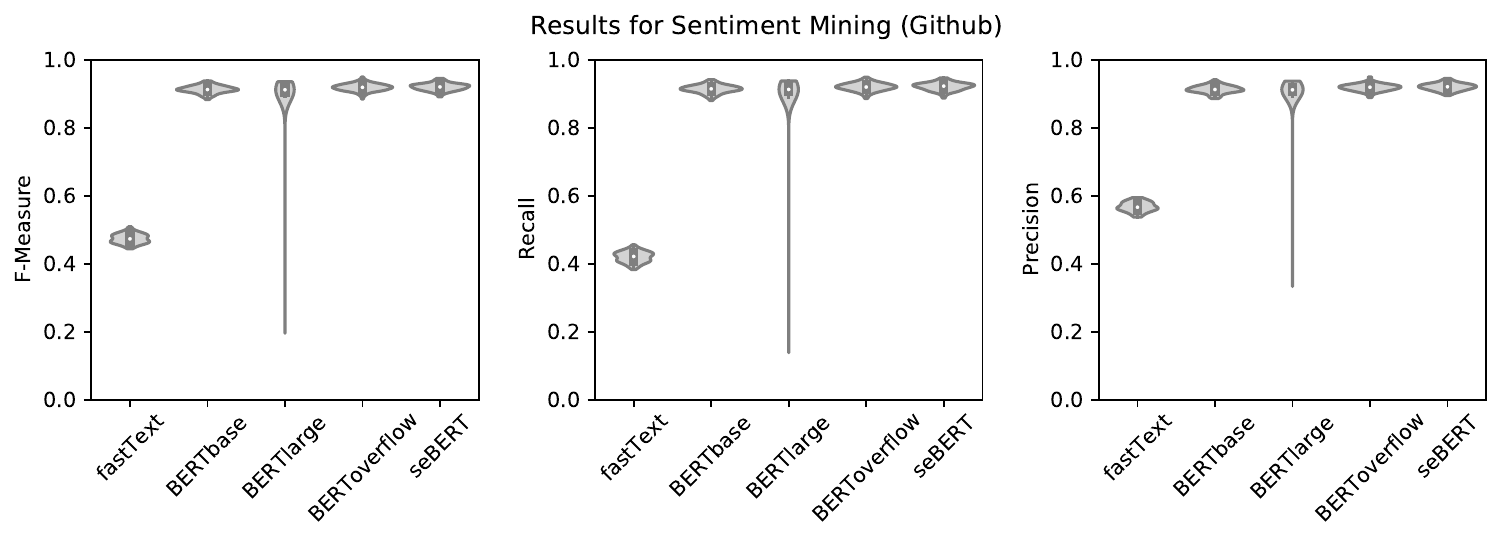}
\caption{Results of the sentiment mining on the GH data.}
\label{fig:results_sentiment_gh}
\end{figure*}

Table~\ref{tbl:fine-tuning} and figures \ref{fig:results_issues}-\ref{fig:results_sentiment_gh} summarize the results of the prediction task comparison. In the following, we look at the results of the individual tasks in detail.

\subsubsection{Issue Type Prediction}

\begin{figure*}
\centering
\includegraphics[width=0.8\textwidth]{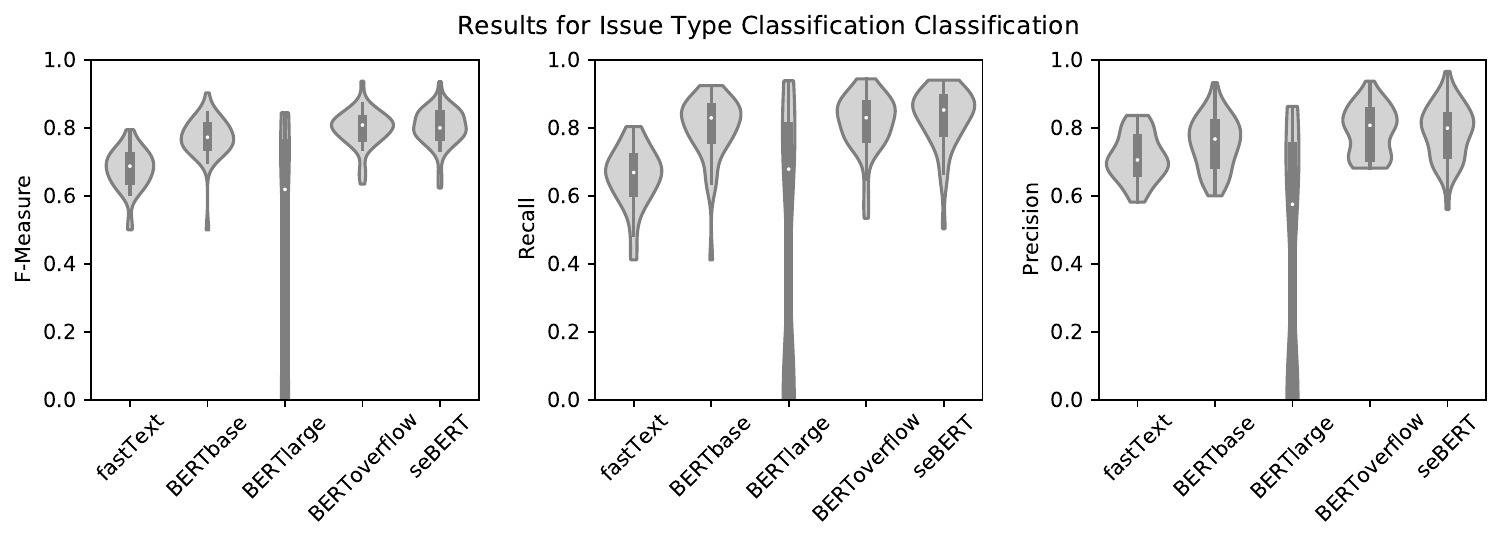}
\caption{Results of the prediction of bug issues.}
\label{fig:results_issues}
\end{figure*}

The results show that seBERT and BERToverflow achieve the best performance for the issue type prediction tasks, outperforming both fastText and the general-domain BERT models. The improvement over fastText is very large with an about 11\% higher \FSCORE{} for the issue type prediction. The violin plot in Figure~\ref{fig:results_issues} indicates that the performance improvement in \FSCORE{} is due to an improvement of both \RECALL{} and \PRECISION{}, which means the models reduced both false positives and false negatives in comparison to fastText. The Bayesian signed rank test determined that this improvement of the SE-specific models over the other models is significant.\footnote{Posterior probabilities determined by the Bayesian signed rank test can be found in the supplemental material.} The difference between seBERT and BERToverflow is not significant The comparison between fastText and BERT\textsubscript{BASE} shows that the general-domain models may be better than smaller text processing models without pre-training on the issue type prediction. BERT\textsubscript{BASE} significantly outperforms fastText on the issue type prediction task. The BERT\textsubscript{LARGE} model from the general domain has severe problems with this use case, i.e., there are several cases where the models completely failed. This happened with none of the other models and may be an indication that the amount of data is too small to fine-tune such a large model from the general domain on a domain-specific corpus.

\subsubsection{Commit Intent Prediction}

\begin{figure*}
\centering
\includegraphics[width=0.8\textwidth]{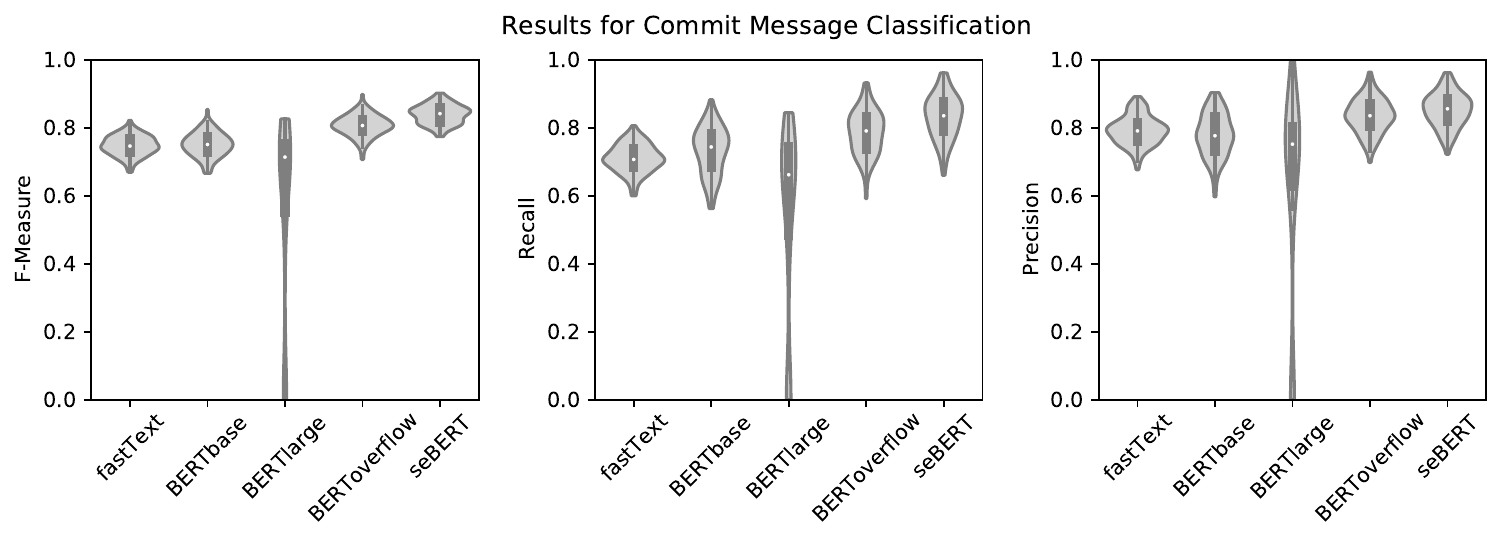}
\caption{Results of the prediction of quality improving commits.}
\label{fig:results_commits}
\end{figure*}

The results for the commit intent prediction are mostly in line with the results for the issue type prediction. seBERT has a 9\% larger \FSCORE{} than the non-SE models, BERToverflow is 6\% better. The violin plots in Figure~\ref{fig:results_commits} also indicate a stable improvement in both \RECALL{} and \PRECISION{} and the Bayesian signed rank test determined that the improvement is significant. In comparison to the issue type prediction, seBERT is significantly better than BERToverflow with an absolute difference of about 3\% in the \FSCORE. Moreover, we note that the general domain BERT models fail to outperform fastText and that BERT\textsubscript{LARGE} has the same stability issues as for the issue type prediction.

\subsubsection{Sentiment Mining}

The results for the sentiment mining show that there are only small differences between the SE domain and general domain BERT models on all three data sets. The absolute performance on the API and GH data is almost equal. The difference on the SO data is slightly larger, where BERToverflow and seBERT outperform the BERT\textsubscript{BASE}  and BERT\textsubscript{LARGE} by about 3\%, with a statistically significant difference. We observe that, same as before, the BERT\textsubscript{LARGE} model is sometimes unstable, likely for the same reasons as above. The difference between the transformer models and fastText is huge for this task, i.e., at least 20\% in \FSCORE{} on all data sets.

\begin{table*}
\centering
\begin{tabular}{llllll}
\toprule
\textbf{Model} & \textbf{Issue Type} & \textbf{Commit Intent} & \textbf{Sentiment (SO)} & \textbf{Sentiment (API)} & \textbf{Sentiment (GH)} \\
\midrule
fastText & 0.69 & 0.75 & 0.44 & 0.37 & 0.47 \\
BERT\textsubscript{BASE} & 0.77 & 0.75 & 0.73 & 0.57 & 0.91 \\
BERT\textsubscript{LARGE} & 0.62 & 0.71 & 0.73 & 0.55 & 0.91 \\
BERToverflow & 0.81 & 0.81 & 0.77 & 0.58 & 0.92 \\
seBERT & 0.80 & 0.84 & 0.76 & 0.57 & 0.92 \\
\bottomrule
\end{tabular}
\caption{Median \FSCORE{} of the models for the fine-tuning tasks. Macro-average over the three classes for the sentiment mining.}
\label{tbl:fine-tuning}
\end{table*}

\section{Discussion}
\label{sec:discussion}

We now discuss our results with respect to our research questions, consider the ethical implications of our work, discuss the limitations and open issues, as well as the threats to the validity. 

\subsection{Interpretation of SE Terminology}

Our results indicate that SE domain models are better at modeling natural language within the SE domain than general domain models. The vocubalaries contain more tokens from the SE domain. Especially the focus of the vocabulary is interesting. Both seBERT and BERToverflow have the names of many tools such as programming languages, libraries, build tools, and operating systems. Such words are mostly missing in the BERT vocabulary. 

That domain specific words replace terms from other domains is not surprising. However, the magnitude of the difference between the vocabularies is larger than we would have expected: only half of the words from BERT are within the BERToverflow vocabulary, for the smaller seBERT vocabulary this drops further to 38\%. Thus, less then half of the commonly used terminology from the general domain is among the commonly used terminology within the SE domain. In comparison, the lexical similarity between English and German is about 60\%~\cite{sim-german-engl}. While this comparison is a bit unfair and likely an overstatement, because the vocabularies require exact matches of word pieces, while the lexical similarity requires only a ``similarity in both form and meaning''~\cite{lexical-similarity}, this highlights how different texts from the SE domain are from the general domain. This also demonstrates that NLP models with a fixed vocabulary should always be retrained from scratch for the SE domain to maximize the performance, instead of basing them on general domain models with additional pre-training steps, as is, e.g., done for BioBERT~\cite{BioBERT} and ClinicalBERT~\cite{ClinicalBERT}. 

The MLM task demonstrates that the better representation of the SE context goes beyond the vocabulary. With SE domain statements, the general domain models often understand that they should suggest a word from the SE domain, but often do not really understand the exact meaning, which leads to unsuitable suggestions. The SE-specific models are much better at understanding the complete context of the missing word and provide suitable suggestion. However, the examples also show that for a SE domain NLP model, data from Stack Overflow alone is not sufficient, as this only represents the domain within a Q\&A context. At least some other aspects of the NLP aspects of the SE domain are not sufficiently captured by the BERToverflow model within our examples. E.g., the lack of issue handling on Stack Overflow translated into a problem with identifying the correct meaning of a missing term. This demonstrates that a larger amount of training data is beneficial and that data should be selected from a diverse range of sources. This also shows that we should be careful, when we use NLP models for tasks, where the text may be different from the pre-training texts. The additional MLM examples confirm what we expected about the SE domain models: their general language understanding is okay, but they are clearly inferior to general domain models for anything beyond the SE domain.

\begin{mdframed}
\textbf{Answer to RQ1:} Transformer models trained with general domain data have trouble understanding SE domain terminology. SE specific models are very good at understanding sentences within the SE context and in general language understanding, but do not perform well in contexts beyond the SE domain. 
\end{mdframed}

\subsection{Impact on Applications}

The results of the fine-tuned predictions show that SE domain transformer models easily are state-of-the-art for the two domain specific NLP tasks we considered, i.e., the issue type prediction and the commit intent prediction. In both tasks, seBERT and BERToverflow clearly outperform the competition and the absolute improvement is huge. Even more importantly, these models are decreasing both false negatives and false positives, i.e., they do not achieve this performance improvement through a trade-off between the errors, which indicates that this is really due to a better understanding of the language. Thus, the high validity of the domain understanding we found through the manual inspection of the vocabularies and the MLM task also translates into a better performance in relevant NLP use cases. This improvement is not explainable by the model size alone. This is demonstrated by the lower performance of the general domain BERT models. These models have the same size as our in domain-models\footnote{Reminder: BERToverflow is a BERT\textsubscript{BASE} model and seBERT is a BERT\textsubscript{LARGE} model.}, but do not show the same improvement over the smaller fastText model. 

Moreover, our results show that the SE domain transformers are not always better when SE data is used: we could not find notable differences between the general domain BERT models and the SE domain models for the sentiment mining tasks. Thus, SE specific pre-training only seems valuable, if the use case is specific to the SE domain and requires the interpretation of text with a SE background, which based on recent research does not seem to be the case for sentiment mining~\cite{Novielli2021}.

The impact of our choice to use a sequence length of only 128, in comparison to the sequence lengths of 512 from the other models is also visible in the fine-tuning. For the short commit messages of the commit intent prediction task, the seBERT model outperforms all other models, including BERToverflow. This is reasonable, because of the seBERT is the overall larger model and commit messages were used for the pre-training. We do not observe such a difference in performance for the issue type prediction task. Here, seBERT and BERToverflow are very similar to each other. This could mean that the \FSCORE{} of 0.8 is the performance ceiling and a better result is not possible without additional data, e.g., more training data or additional information about the issues. Alternatively, seBERT is penalized by the short sequence length of 128, which gives BERToverflow an advantage for longer issues. This could compensate the difference in model size and data for pre-training. A subgroup analysis of the predictions for issues with less than or equal to 128 tokens and longer issues did not provide conclusive evidence for this. Instead, we observed an instance of Simpson's paradox~\cite{Blyth1972}, i.e., seBERT outperformed BERToverflow on both subgroups, even though the overall performance is about equal. While we could consider this as an indication that the sequence length was not problematic, we cannot provide a definitive answer to this question without extending seBERT to a sequence length of 512 tokens. However, such an extension is, at this time, from our perspective not warranted due the required energy consumption (see Section~\ref{sec:ethics}) and we believe that the computational effort for this should only be spent if there is a clear expected advantage of considering longer sequences.

The difference between the BERT\textsubscript{BASE} and BERT\textsubscript{LARGE} model further shows that general-domain models -- especially large general domain models -- should be used with caution: often, this model converged to a trivial result where all instances were classified as negative. The logs of the training showed that this happened when the fine-tuned model was still very bad after the first epoch. In the subsequent epochs, the optimization found that a trivial model was actually better in terms of accuracy and cross-entropy than the first attempts. Unfortunately, we cannot determine the exact reason for this. However, we believe that this may be due to the combination of a small data set for fine-tuning and a lack of SE domain understanding of the general model. As a result, the training did not converge towards a reasonable solution within one epoch. Subsequently, the optimizer got stuck in the local optimum of the trivial model.\footnote{We cannot rule out that other training parameters (e.g., batch size, learning rate), would yield better results with BERT\textsubscript{LARGE} or any other of our models. However, this does not affect our conclusions, as we discuss in the threats to validity (see Section~\ref{sec:threats}).} That this did not happen with the equally large seBERT provides another indication that the SE specific pre-training, in fact, the opposite happened: since the pre-training already captured the domain very well, seBERT usually achieved the optimal result within one, at most two epochs, regardless of the model size.

Together with the results from Tabassum\etal~\cite{tabassum-etal-2020-code} for the NER task that showed that the BERToverflow model was also better than a BERT\textsubscript{BASE} model, we have a clear answer for RQ2. 

\begin{mdframed}
\textbf{Answer to RQ2:} Transformer models pre-trained with SE domain data consistently outperform other models on SE use cases and should be considered as the state-of-the-art. Pre-trained models from the general domain should be used with care, especially if only a small amount of data is available for the fine-tuning. However, we also found that SE domain models perform similar to general domain models for sentiment mining on SE data, i.e., a use case that does not require SE specific knowledge. 
\end{mdframed}

\subsection{Ethical Considerations}
\label{sec:ethics}

Large deep learning models for NLP, like the transformer models we consider within this work, are associated with several ethical challenges, as is, e.g., highlighted in the famous stochastic parrots paper by Bender\etal~\cite{Bender2021}. Due to impact of Bender\etal~\cite{Bender2021} and the direct relation to BERT models, we structure our consideration of ethical aspects following the four major ethical concerns highlighted in their work. 

The first aspect highlighted by Bender\etal~\cite{Bender2021} is the energy consumption that large transformer models require. We actively considered methods to reduce the training time, e.g., by using a version of the BERT pre-training that was optimized for the hardware we were using and by restricting the sequence length to 128 tokens. Based on the energy consumption of the system we used for training that requires between 2.5 kW and 3.5 kW when utilized fully, we estimate that we required between 270 kWh and 378 kWh for the 4.5 days of pre-training. Under load the cooling of the data center, where the compute nodes are located, has a Power Usage Effectiveness (PUE) of about 1.23. Additional overhead regarding storage and network has also to be taken into account, so we are calculating with an overhead of approximately 30\%. Hence, we estimate that we consumed at most $378~\text{kWh} \cdot 1.30 = 491.4~\text{kWh}$. Based on the 366 g CO\textsubscript{2} that is generated per kWh in 2020 in Germany~\cite{Icha2021}, this means the pre-training produced up to 180 kg of CO\textsubscript{2}. This is roughly the amount of the CO\textsubscript{2} for one tank filling of an average family car, which we believe is not unreasonable from an ecological perspective for a one-time effort. In comparison, Strubell\etal~\cite{Strubell2019} estimate that they required about 1507 kWh for the training of a BERT\textsubscript{BASE} model. This means that we could train the seBERT model with only one third of the environmental impact than a normal BERT\textsubscript{BASE} model, even though we use a BERT\textsubscript{LARGE} architecture and 119.7 Gigabyte instead of 16 Gigabyte of textual data for training. 

The second aspect highlighted by Bender\etal~\cite{Bender2021} is the quality of the training data. The authors highlight that the size of the training data does not guarantee diversity, data collected from the past can, by definition, capture how language evolves, and that there is a risk that models, therefore, capture and enforce existing biases. We did not systematically evaluate BERToverflow and seBERT for such biases. One aspect we found regardless is that ``women'' was not in the dictionary of seBERT. Thus, we can be quite sure, even without an in depth consideration that there is at least a severe gender bias within the data, and, consequently, within seBERT. Since BERToverflow was trained on less, but similar data, we believe that there should be a similar gender bias. We cannot comment on other biases, e.g., racial bias or similar. We also note that we cannot exclude that men is only included in the seBERT vocabulary because it is polysemous (man page). Tasks like NER or classification of bugs should not be affected by such biases and can safely be used. Tasks where bias is be relevant, e.g., sentiment mining of developers, the development of chat bots or automated answering of SE questions should only be conducted after a detailed consideration of such ethical considerations. For tasks where the impact of bias is unclear, e.g., the summarizing existing texts within the SE domain, could possibly be used but we still recommend to at least conduct a basic check for such biases. 

The third aspect that Bender\etal~\cite{Bender2021} consider is that time may be better spent than on the exploration of ever larger language models. When we transfer this to our work, this means that SE researchers should not invest too much effort into the development of NLP models for the SE domain, but rather focus on the SE data and use cases. For us, this means that we should only provide SE domain models, when machine-learning driven NLP research has major advances, instead of, e.g., trying to directly find new transformer architectures for SE benchmarks with SE data. Currently, it seems like the SE community is already using such an approach, as there are only few domain specific models and they are all re-using architectures that were developed by researchers for the general domain (see Section~\ref{sec:related-work}).

The fourth issue raised by Bender\etal~\cite{Bender2021} is that, in the end, such large NLP models are nothing more than \textit{stochastic parrots}, i.e., models that repeat what they have seen in the data, with some random component. This is due to our lack of understanding about the internal structure and reasoning within models with millions of parameters. This is also a property of SE domain models  that should be respected for any later use of these models in an ethical way. For example, we found that the SE domain models are very good at suggesting technologies. When we use the sentence ``You can use [MASK] for code coverage in java.'', seBERT and BERToverflow predict tools like Cobertura, gcov, Emma, or Jacoco. We found that this works with different languages and different kinds of tools. However, using the models within a chat bot or Q\&A system to recommend suitable tools means that the past is encoded and developers of new tools would not have a chance, unless the model is retrained. In comparison, humans learn continuously about new tools, i.e., there would not be such an ethical problem. The consequence of this issue is similar to the impact of potential biases: we recommend to be careful when using the NLP models to generate responses to actual queries, unless it was determined that the possible responses are carefully validated for the given use case. 

\subsection{Open Issues}
\label{sec:limitations}

Our work demonstrates that NLP models pre-trained with SE domain data are useful. However, there are also limitations to the understanding of transformers that we established. Since we used BERT both as general domain reference model, as well as the architecture for our models, it is unclear how the results generalize to other transformer models. While the difference to models that have a similar size like RoBERTa~\cite{liu2019roberta} should be relatively small, it is unclear if extremely large models like GPT-3~\cite{brown2020language} may be able to correctly understand the meaning of texts both in the SE and general domain, same as human experts. While there is no reason that this should be the case, there is also no strong argument against this. However, currently the scale of these models is beyond almost anyone, except the largest labs and companies of the globe~\cite{Hellendoorn2021}. Therefore, even if this were the case, such improvements could currently only be harnessed by a small elite. Thus, we believe that for the majority of SE researchers and vendors who may consider building NLP capabilities into their tools, models like BERT are a more realistic, current alternative. 

A related limitation is the impact of the context length on the results. Within this work, we work with a maximum context length of 512 tokens (BERT, BERToverflow) and 128 tokens (seBERT). While we argue that most SE texts in our data are shorter anyways, this is also due to the type of text we consider. If we were to, e.g., consider README files instead, we would likely have many examples of longer texts. Thus, while our results show that even the relatively short context of 128 tokens is sufficient and actually leads to the best results in both domain-dependent fine-tuning examples we consider, this may not generalize to NLP tasks on longer inputs. Especially generative tasks, like summarizing long documents, may benefit from a longer context that is sufficient to capture the meaning of the whole text at once. In case studies find that models with shorter contexts, like BERT, do not generate suitable results for longer documents, other transformer models, like Big Bird~\cite{NEURIPS2020_c8512d14}, which achieves a sequence length of 4096 tokens, could be used. 

Another limitation is a corollary from our statement regarding longer documents: while our corpus is already relatively diverse, with Stack Overflow, issues, and commit messages, this still does not capture the whole SE domain and, most notably, lacks examples with longer texts. Thus, even if we wanted to train a model with a longer sequence length, this could only make a difference on a small fraction of this type of SE data and the available data would likely not be sufficient to correctly model longer contextual relationships. Unfortunately, there is neither a suitable data set that could be exploited for pre-training, nor is there a benchmark task for NLP within the SE domain that requires longer texts. Thus, to advance NLP for the SE domain for tasks with long sequences, our community would first need to solve the associated data challenge, both with respect to data for pre-training, as well as through a curated data set that is suitable for the benchmarking of a fine-tuned application. These limitations are not only restricted to length, but also to tasks that actively exploit the capability of BERT models to compare two sentences by exploiting the semantics of the [SEP] token. Thus, further fine-tuning tasks that evaluate the utility of BERT for such problems, e.g., the identification of related Stack Overflow posts~\cite{Xu2018}, still need to be considered. Additionally, similar fine-tuning tasks to the ones we considered to evaluate the generalization of the results to similar tasks, e.g., app review classification~\cite{Maalej2016}, which is similar to sentiment mining. 

Finally, we already highlighted the lack of an evaluation of the SE models from an ethics perspective. While this was not within the scope of our work, future work must deal with potential biases in SE domain models, unless their usage is restricted to few and possibly uncritical use cases, as is, e.g., the case in our fine-tuned examples. We note that this does not only affect seBERT and BERToverflow, but also models like CodeBERT~\cite{feng-etal-2020-codebert}, in case this is used to generate texts, e.g., automated documentation generation. From our perspective, this is a precursor of any type of generative NLP application, i.e. NLP models that actively generate texts, e.g., to answer questions like ``what does this code do'' or ``summarize this README file'', but also for any application like sentiment mining, which is known to encode biases~\cite{kiritchenko-mohammad-2018-examining}. 

\subsection{Threats to Validity}
\label{sec:threats}

We report the threats to the validity of our work following the classification by \cite{Cook1979} suggested for software engineering by \cite{Wohlin2012}. Additionally, we discuss the reliability as suggested by \cite{Runeson2009}. 

\subsubsection{Construct Validity}

The construct of our validation of NLP models may be unsuitable. The direct comparison of WordPiece vocabularies neglects to account for the internal structure of the models. For example, the internal structure could achieve that the combination of tokens ``wo\#\#men'' is the same as having the token ``women'' directly in the vocabulary. We address this issue by not only considering the overlap and tokenization, but also through a qualitative analysis of the non-overlapping words. Our data indicates that these are mostly domain specific terms, which makes sense given the training data and also means that an effect where these words are known by the model, regardless of them missing in the vocabulary. Similarly, our study of the contextual interpretation of sentences and words through the MLM task may be unsuitable. Our lack of consideration of different parameters for training (batch sizes, learning rates, etc.) may lead to sub-optimal models after fine-tuning, which could affect our conclusions. 

\subsubsection{Internal Validity}

While we have good reason to believe that the differences between the models we observe are due to the difference in the data used for pre-training the models, we cannot rule out that there may be other reasons for these differences, due to the black box nature of the models. Most threats to the internal validity should be mitigated by our construct: a pure analysis of the vocabulary may show artificial differences, but this would not explain the differences we observe with the MLM and fine-tuning tasks. And while few random differences for the MLM and fine-tuning may be explainable by alternative hypotheses, we observe clear patterns that match our expectations, including the lack of big differences between the models when general language understanding is required. We also did not conduct an extensive hyperparameter search for the fine-tuning of the models. This means we potentially underestimate the models performance, such that differences between the models could potentially increase or decrease, with other parameters, e.g., learning rates. However, since all models use similar architectures (BERT), it is likely that they work best with similar hyperparameters for the same problem. Thus, even if our hyperparameters should be suboptimal, it is likely that the differences we observe would be preserved with different hyperparameters. Consequently, we believe that we mitigated most notable threats to the internal validity, other than the limitations of our study we acknowledge in Section~\ref{sec:limitations}. 

There is one notable threat to the internal validity of our fine-tuning results, due to the way seBERT was created. The corpus used for pre-training included commit messages from Github, data from Stack Overflow, and Jira issues, some of which may be part of the test data we use for fine-tuning tasks. Since the pre-training was self-supervised based on MLM and NSP, no information about labels was part of the creation of the models. Thus, there cannot be an information leak that affects our results. However, we cannot rule out the possibility that the fine-tuning works better with data seen during the pre-training, because it is more likely that the language structure is already known by the model. However, we believe that the threat is relatively small, due to the following reasons. First, having seen the data could also be a problem as well: beyond language understanding, having seen data without the labels before can only help with memorization, not with generalization. Consequently, results on test data could actually be worse due to this. Second, while some data may have been during the pre-training, we also have fine-tuning tasks, where none (API sentiments), or only little data was part of the pre-training (prediction of bug issues\footnote{The Jira Issues only contain data until 2014 and the difference in performance of predictions is only later data is considered is small~\cite{Herbold2020}}). Third, BERToverflow used different data and can only have such an overlap for the sentiment mining on SO task, but there is no indication that this is an advantage for seBERT for the other tasks. Fourth, note that the data we used for fine-tuning is tiny (a couple of Megabytes) compared to the data used for pre-training (e.g., 2.4 Terabyte for seBERT). Thus, it is unlikely that the pre-training was significantly influenced by this data, making advantages unlikely.

\subsubsection{Conclusion Validity}

The statistical tests we used for the comparison of the fine-tuned models were suitable for the data and there is no threat to the conclusion validity of our study. 

\subsubsection{External Validity}

Since our consideration of fine-tuning only considers five classification tasks, we cannot with certainty conclude that the SE domain models are also better than the general domain models for other SE tasks, e.g., summarizing content or comparing sequences. However, since our results indicate that SE-specific models are better at capturing SE specific aspects and because prior work also found a similar results regarding the NER task~\cite{tabassum-etal-2020-code}, we believe that this is unlikely. Moreover, as already discussed in Section~\ref{sec:limitations}, our results may not generalize to extremely large models, as they may be able to capture multiple domains correctly, due to their size. We note that our pre-training and evaluations are limited to data collected from open source projects. While many open source projects are developed professionally, there may still be issues due to unknown in-house terminology when these models are used within a proprietary context that requires such knowledge. Our results do not allow for conclusions regarding the question if fine-tuning of open source models would be sufficient for such use cases.

\subsubsection{Reliability}

The definition of the positive, neutral, and negative sentences for the assessment of the contextual embeddings may influence our results, i.e., other researchers would almost certainly not have selected exactly the same sentences as us and, therefore, may have a different view on the capability of the models with respect to understanding the context of statements. However, while other sentences could always lead to differences, we note that we did not conduct any cherry picking to show the most differences, but rather first defined the sentences and then applied the models. Moreover, our playground provides anyone with the ability to evaluate the differences in the predictions of masked words between the models.\footnote{\url{https://smartshark2.informatik.uni-goettingen.de/sebert/index.html}} Additionally, we augmented the analysis on the curated corpus with a pure empirical analysis, in which we randomly sampled sentences with words with special meaning in the SE domain and then evaluated the ability of the models to predict the missing words. While this does not give us insights into the inner workings of the models, these empirical results provide further evidence that our results are not biased by our selection.

\section{Conclusion}
\label{sec:conclusion}

Within our work, we find that NLP for SE can benefit from transformer models pre-trained with SE data. While this was already known for applications that include source code, there was only little evidence for other natural language tasks. Through our work we not only explored the differences in the expected performance of applications, but also the tried to understand if the models really have a better understanding of the SE domain. For this, we manually compared the vocabularies of general domain and SE domain BERT models and compared how these models completed sentences with masked words. This analysis showed that while general domain models have a rough but imprecise understanding of the SE domain, the SE models are more precise. This improved understanding also led to significantly better results in fine-tuned within the SE domain, but not in a general task like sentiment mining applied to SE data. In conclusion, we recommend to ensure that a large amount of SE data was used for pre-training large NLP models, when these models are used for SE tasks.

\ifCLASSOPTIONcompsoc
  \section*{Acknowledgments}
\else
  \section*{Acknowledgment}
\fi

The authors would like to thank the GWDG for their support regarding the usage of the GPU resources required for this article.

\bibliographystyle{IEEEtran}
\bibliography{literature.bib}

%

\begin{IEEEbiography}{Julian von der Mosel}
Julian von der Mosel is currently working as a data engineer and received his B.Sc. degree from the University of Goettingen.
\end{IEEEbiography}

\begin{IEEEbiography}{Alexander Trautsch}
Alexander Trautsch is a PhD candidate at the University of Goettingen.
\end{IEEEbiography}


\begin{IEEEbiography}{Steffen Herbold}
Steffen Herbold is professor for Methods and Applications of Machine learning at the TU Clausthal. 
\end{IEEEbiography}


\vfill


\end{document}